\documentclass[aps,prb,floatfix,twocolumn,superscriptaddress,showpacs,amsmath,amssymb]{revtex4-1}
\usepackage{graphicx}
\usepackage{epstopdf}
\usepackage{epsfig} 
\allowdisplaybreaks

\newcommand{\be}{\begin{equation}}
\newcommand{\ee}{\end{equation}}
\newcommand{\bea}{\begin{eqnarray}}
\newcommand{\eea}{\end{eqnarray}}
\newcommand{\ba}{\begin{eqnarray*}}
\newcommand{\ea}{\end{eqnarray*}}
\newcommand{\dagga}{{\phantom{\dagger}}}

\newcommand{\bq}{\mathbf{q}}

\newcommand{\bx}{\mathbf{x}}

\newcommand{\Ima}{{\Im m}}
\newcommand{\Rea}{{\Re e}}
\newcommand{\dis}{\displaystyle}

\newcommand{\up}{\uparrow}
\newcommand{\down}{\downarrow}
\newcommand{\fract}[2]{\frac{\dis \;#1\;}{\dis \;#2\;}}
\newcommand{\Tr}{\mathrm{Tr}}
\newcommand{\eqn}[1]{(\ref{#1})}
\newcommand{\ket}[1]{\mid\! #1\rangle}
\newcommand{\bra}[1]{\langle #1\!\mid}

\newcommand{\bw}{\begin{widetext}}
\newcommand{\ew}{\end{widetext}}

\newcommand{\cW}{\mathcal{W}}
\newcommand{\hPhi}{\hat{\Phi}}

\newcommand{\bphi}{\boldsymbol{\phi}}
\newcommand{\bsigma}{\boldsymbol{\sigma}}
\newcommand{\bo}[1]{\boldsymbol{#1}}
\newcommand{\z}{{(0)}}%

\begin{document}

\title{Quantum fluctuations beyond the Gutzwiller approximation}
\author{Michele Fabrizio} 
\affiliation{International School for
  Advanced Studies (SISSA), Via Bonomea
  265, I-34136 Trieste, Italy}

\date{\today} 


\begin{abstract}
We present a simple scheme to evaluate linear response functions including quantum fluctuation corrections on top of the Gutzwiller approximation. The method is derived for a generic multi-band lattice Hamiltonian without any assumption about the dynamics of the variational correlation parameters that define the Gutzwiller wavefunction,  and which thus behave as genuine dynamical degrees of freedom that add on those of the variational uncorrelated Slater determinant. \\
We apply the method to the standard half-filled single-band Hubbard model. We are able to recover known results, but, as by-product, we also obtain few novel ones. In particular, we show that quantum fluctuations can reproduce almost quantitatively the behaviour of the uniform magnetic susceptibility uncovered by dynamical mean field theory, which, though enhanced by correlations,  is found to be smooth across the paramagnetic Mott transition. By contrast, the simple Gutzwiller approximation predicts that susceptibility to diverge at the transition.  

\end{abstract}
\maketitle

\section{Introduction}
The Gutzwiller approximation\cite{Gutzwiller-1,Gutzwiller-2} is likely the simplest tool to deal with strong correlations in lattice models of interacting electrons. It consists in a recipe for approximate analytical expressions of expectation values in a class of wavefunctions, named Gutzwiller wavefunctions,  of the form  
\be
\ket{\Psi} = \prod_i\,\mathcal{P}(i)\ket{\Psi_0}\,,
\label{intro-Psi}
\ee
where $|\Psi_0\rangle$ is a variational Slater determinant, and 
$\mathcal{P}(i)$ a linear operator that acts on the local Hilbert space at site $i$ and depends on a set of variational parameters. \\
Curiously, the Gutzwiller approximation often provides physically more sound results than a direct evaluation of expectation values in wavefunctions like Eq.~\eqn{intro-Psi}. For instance, the numerical optimisation on a finite-dimensional lattice of a variational Gutzwiller wavefunction for a single-band half-filled Hubbard model  never stabilises a genuine Mott insulating phase\cite{Shiba1987,Capello2005}, i.e. an insulator that does not break any symmetry, which intuitively is to be 
expected beyond a critical strength of the on-site repulsion. By contrast, the Gutzwiller approximation is instead able to describe such a genuine Mott transition\cite{Brinkman&Rice}. The explanation of this strange outcome relies on the following observations. The first is that, in order to describe a genuine Mott insulator, one needs to add to the Gutzwiller wavefunction, Eq.~\eqn{intro-Psi}, long range density-density Jastrow factors\cite{Capello2005}.  However, the effect of such Jastrow factors disappears in lattices with coordination number $z\to\infty$, therefore, only in that limit, wavefunctions like Eq.~\eqn{intro-Psi} can faithfully describe Mott insulators. Moreover, right in that limit of $z\to\infty$, the Gutzwiller approximation provides the exact expression of expectation values\cite{Metzner1989,BW&G1998}. Therefore the Gutzwiller approximation should better be regarded as a recipe to evaluate approximate expectation values in Gutzwiller-Jastrow wavefunctions, which becomes exact when the coordination number tends to infinity, rather than in Gutzwiller-only wavefunctions. In other words, the Gutzwiller approximation applied on a lattice with finite $z$ is just the variational counterpart of dynamical mean field theory (DMFT)\cite{ReviewDMFT} applied on that same lattice. \\
Recently, several attempts to include the Gutzwiller approximation inside DFT electronic structure codes have been performed with quite encouraging outcomes\cite{Fang2008,Ho2008,Fang2009,Fang2010,Fang2011,Ho2012,Weber2012,Lanata2013,Lanata2014,Borghi2014,Fang2015,NicolaPRX}.  In this perspective, it might be useful to have at disposal a simple and flexible method to calculate linear response functions within the Gutzwiller approximation, in view of an extension 
of the so-called linear response TDDFT\cite{Gross1984,Gross1996} to the case when DFT is combined with the Gutzwiller approximation. \\
There are already several works dealing with linear response in the Gutzwiller approximation, most of which 
limited to the single-band Hubbard model\cite{VollhardtRMP,LorenzanaPRL,Lorenzana-PRB2003,LorenzanaPRB2004,SchiroFabrizioPRB,Lorenzana&Capone2013}. Extensions to multi-band models have been attempted 
\cite{Seibold2011a,Seibold2011b}, though under an assumption about the dynamics of the variational parameters that determine the linear operators $\mathcal{P}(i)$ in Eq.~\eqn{intro-Psi}. \\
Here we shall instead present a very simple and general method to evaluate linear response functions within 
the Gutzwiller approximation without any preliminary assumption. The method is essentially an extension of the time-dependent Gutzwiller approximation of Ref.~\onlinecite{SchiroFabrizioPRL} to a generic multi-band Hamiltonian, where the dynamics of the linear operators $\mathcal{P}(i)$ and of the Slater determinant 
$|\Psi_0\rangle$, see Eq.~\eqn{intro-Psi}, are treated on equal footing. Linearisation of the equations of motion around the stationary solution, which is the equilibrium state, thus allows calculating linear response functions. \\
We note that the results of the Gutzwiller approximation at equilibrium coincide with the saddle point solution of the slave-boson theory in the path-integral formulation\cite{Kotliar&Ruckenstein}, which,  in multi-band models, corresponds to the so-called rotationally invariant slave boson formalism (RISB)\cite{RISB}. Our present results in the linear response regime can therefore be considered equivalent to the quantum fluctuations corrections above the RISB saddle-point solution. We preferred here to derive such corrections to the action directly from the time-dependent Gutzwiller approximation rather than from the RISB theory, since the former is at least a well controlled variational scheme in lattices with infinite coordination number. However, both the notations as well as the language we shall use are actually closely related to RISB theory.  \\

The paper is organised as follows. In Sec.~\ref{The Gutzwiller approximation in brief} we briefly present the time-dependent Gutzwiller approximation, with some additional technical details postponed to the Appendix. In Sec.~\ref{Fluctuations above the saddle point solution} we linearise the equations of motion around the stationary solution and derive an effective action for the fluctuations in the harmonic approximation. In Sec.~\ref{Application to the half-filled Hubbard model} we apply the method to the single-band half-filled Hubbard model, which allows a comparison with already existing results. 
Section~\ref{Conclusions} is devoted to concluding remarks.

\section{The Gutzwiller approximation in brief}
\label{The Gutzwiller approximation in brief}
Besides the original works\cite{Gutzwiller-1,Gutzwiller-2} where M. Gutzwiller introduced a novel class of  variational wavefunctions as well as an approximate scheme to compute expectation values, after him called Gutzwiller wavefunctions and approximation, and the subsequent demonstration that such an approximation becomes exact in the limit of infinite-coordination lattices\cite{Metzner1989,BW&G1998}, 
there are by now many articles where the Gutzwiller approximation is described in detail. Here we shall follow Ref.~\onlinecite{Fabrizio_review} and use its same notations. \\
The time-dependent Gutzwiller wavefunction is defined through
\cite{LorenzanaPRL,SchiroFabrizioPRL,Fabrizio_review}
\be
\ket{\Psi(t)} = \prod_i\, \mathcal{P}(i,t)\ket{\Psi_0(t)},\label{GW}
\ee
which is the analogous of Eq.~\eqn{intro-Psi} where now 
$\ket{\!\Psi_0(t)}$ is a time-dependent variational Slater determinant, and $\mathcal{P}(i,t)$ 
linear operators on the local Hilbert space that depend on time-dependent variational parameters. For sake of simplicity, we shall not include in our analysis BCS wavefunctions nor operators $\mathcal{P}(i,t)$ that are charge non-conserving. The extension to those cases is simple, though notations get more involved.\\
Suppose that the Hamiltonian is written in terms of fermionic operators $c^\dagga_{i\alpha}$ 
and $c^\dagger_{i\alpha}$, $\alpha=1,\dots,2M$, that correspond to annihilating or creating a fermion at site $i$ in a chosen basis of Wannier functions $\phi_{i\,\alpha}(\bx,t)$, where $\alpha$ indicates both spin and orbital indices. Let us imagine a $U(2M)$ unitary transformation 
\be
\mathcal{W}(i,t) = \exp\bigg(
i\sum_{\alpha\beta}\,K_{\alpha\beta}(i,t)\,c^\dagger_{i\alpha}\,c^\dagga_{i\beta}\bigg)\,,
\label{qq-W}
\ee
with $K_{\alpha\beta}(i,t) = K_{\beta \alpha}(i,t)^*$, which maps $c_{i\,\alpha}$ into a new basis set $d_{i\,\alpha}$ of single particle operators 
\be
d^\dagga_{i\,\alpha} = \mathcal{W}(i,t)^\dagger\,c^\dagga_{i\,\alpha}\,
\mathcal{W}(i,t) = \sum_\beta\,U_{\alpha\beta}(i,t)\,c^\dagga_{i\,\beta}.
\label{f-c}
\ee
Evidently, if we consider the \textit{gauge} transformation  
\bea
\mathcal{P}(i,t) &\to& \mathcal{P}(i,t)\,\mathcal{W}(i,t)^\dagger\,,
\label{W-P}\\
\ket{\Psi_0(t)} &\to& \prod_i\,\mathcal{W}(i,t)\ket{\Psi_0(t)}\,,\label{W-Psi}
\eea
the Gutzwiller wavefunction $|\Psi(t)\rangle$ in \eqn{GW} stays invariant and the transformed 
$|\Psi_0(t)\rangle$ remains a Slater determinant. Such gauge invariance, analogous to that of the RISB theory\cite{RISB}, repeatedly appears in the calculations that follow. \\

The most general $\mathcal{P}(i,t)$ can be written\cite{mio-dimero,Fabrizio_review} as 
\bea
\mathcal{P}(i,t) &=& \sum_{n\bar{m}}\, \lambda_{n\bar{m}}(i,t)\,\ket{n;i}\bra{\bar{m};i},
\eea
where $n$ and $\bar{m}$ can be chosen to belong to the local basis 
of Fock states built with the operators $c^\dagga_{i\,\alpha}$. Alternatively, one can use a \textit{mixed-basis} representation where $n$ labels Fock states in the original basis $c^\dagga_{i\,\alpha}$, 
and $\bar{m}$ Fock states in a different basis\cite{NicolaPRB2008}, e.g. the basis of the operators 
$d^\dagga_{i\alpha}$ in Eq.~\eqn{f-c}, which is also used to built the Slater determinant $|\Psi_0(t)\rangle$. 
We define the uncorrelated local probability distribution $\hat{P}_0(i,t)$, which is positive definite, by its matrix elements
\be
P_{0\,\bar{n}\bar{m}}(i,t) = \big\langle\Psi_0(t)\Big| \, \ket{\bar{m};i}\bra{\bar{n};i}\,\Big|\Psi_0(t)\big\rangle,
\ee
as well as the Gutzwiller variational matrix 
\be
\hat{\Phi}(i,t) \equiv \hat{\lambda}(i,t)\,\sqrt{\,\hat{P}_0(i,t)\;}\;,\label{def:Phi}
\ee
with matrix elements $\Phi_{n\bar{m}}(i,t)$. Expectation values of local and non-local operators 
in the Gutzwiller wavefunction \eqn{GW} can be calculated explicitly in infinite coordination lattices if one 
imposes the following two constraints at any time\cite{BW&G1998,Fabrizio_review}:
\begin{align}
\Tr\Big(\hat{\Phi}(i,t)^\dagger\,\hat{\Phi}(i,t)\Big) &= 1,\label{1}\\
\Tr\Big(\hat{\Phi}(i,t)^\dagger\,\hat{\Phi}(i,t)\,\hat{c}^\dagger_{i\,\alpha}
\hat{c}^\dagga_{i\,\beta}\Big) &\equiv n_{\alpha\beta}(i,t)\nonumber\\
=  &\;\bra{\Psi_0(t)} c^\dagger_{i\,\alpha}c^\dagga_{i\,\beta} \ket{\Psi_0(t)}, \label{2}
\end{align}
where the fermionic operators within the spur must be regarded as their matrix representation in the local 
Fock space. The second constraint Eq.~\eqn{2} plays the role of a 
\textit{gauge-fixing} condition, exactly as in the RISB model\cite{RISB}.\\
Another important ingredient is the wavefunction renormalisation matrix $\hat{R}(i,t)$ with elements 
$R_{\alpha \beta}(i,t)$, defined by solving the set of equations
\begin{align}
& \bra{\Psi_0(t)} c^\dagger_{i\gamma}\,\mathcal{P}(i,t)^\dagger \, c^\dagga_{i\alpha}\,
\mathcal{P}(i,t) \ket{\Psi_0(t)} \nonumber\\
& \qquad \qquad \qquad \qquad = \sum_\beta n_{\gamma\beta}(i,t) \;R_{\alpha \beta}(i,t)^\dagga \,,\label{definizione-R}
\end{align}
where the left hand side can be straightforwardly evaluated by the Wick's theorem. 
As shown in the Appendix \ref{app-1-1}, the solution of the above equation reads
\be
\hat{R}(i,t) = \hat{Q}(i,t)\,\hat{S}(i,t)\,,\label{R}
\ee
where $\hat{Q}(i,t)$ has matrix elements
\be
Q_{\alpha\beta}(i,t) = \Tr\Big(\hPhi(i,t)^\dagger\,\hat{c}^\dagga_{i\alpha}\,\hPhi(i,t)\,\hat{c}^\dagger_{i\beta}\Big)\,,
\label{Q-def}
\ee
and the hermitian matrix $\hat{S}(i,t)$ is defined through
\be
4\hat{S}(i,t)^{-2} = 1 - \hat{\Delta}(i,t)^2
\,\label{S-def}
\ee
where the matrix elements of  $\hat{\Delta}(i,t)$ are 
\be
\Delta_{\alpha\beta}(i,t) = \Tr\bigg(\hPhi(i,t)^\dagger\,\hPhi(i,t)\,
\Big[\,\hat{c}_{i\alpha}^\dagga\,,\,\hat{c}_{i\beta}^\dagger\,\Big]\bigg)\,.\label{Delta-def}
\ee
The meaning of $\hat{R}(i,t)$ is that the action of the annihilation operator $c^\dagga_{i\alpha}$ on the 
Gutzwiller wavefunction is equivalent to the action of the operator
\be
\mathcal{P}(i,t)^\dagger\,\mathbf{c}_i^\dagga\,\mathcal{P}(i,t) \to \hat{R}(i,t)\,\mathbf{c}_i\, 
\label{R-transform}
\ee
on the Slater 
determinant $|\Psi_0(t)\rangle$, where $\mathbf{c}_i$ is a spinor with components $c_{i\alpha}$. 
One can readily show that under the gauge transformation Eq.~\eqn{W-P},
\be
\hat{R}(i,t) \to \hat{R}(i,t)^W=\hat{R}(i,t)\,\hat{U}(i,t)^\dagger\,,\label{W-R}
\ee
where $\hat{U}(i,t)$ has the matrix elements $U_{\alpha\beta}(i,t)$ of Eq.~\eqn{f-c}, so that Eq.~\eqn{R-transform} transforms into 
\[
\mathcal{W}(i,t)\mathcal{P}(i,t)^\dagger\,\mathbf{c}_i^\dagga\,\mathcal{P}(i,t) \mathcal{W}(i,t)^\dagger 
\to\; \hat{R}(i,t)^W\,\mathbf{d}_i\,.
\]
Since we have complete freedom in choosing $\mathcal{W}(i,t)$, a convenient choice is the unitary transformation that diagonalises the local single-particle density matrix, in which case the operators 
$d^\dagga_{i\alpha}$ are associated to the natural orbitals and satisfy 
\be
\Tr\Big(\hat{\Phi}(i,t)^\dagger\,\hat{\Phi}(i,t)\,\hat{d}^\dagger_{i\,\alpha}
\hat{d}^\dagga_{i\,\beta}\Big) = \delta_{\alpha\beta}\,n_\alpha(i,t)\,,\label{natural}
\ee
while the matrix elements of $\hat{R}(i,t)^W$ acquire the simple expression
\bea
R_{\alpha\beta}(i,t)^W &=& \fract{\;\Tr\Big(\hPhi(i,t)^\dagger \hat{c}^\dagga_{i\alpha} \hPhi(i,t) \,
\hat{d}^\dagger_{i\beta}\Big)\;}{\;\sqrt{n_\beta(i,t)\Big(1-n_\beta(i,t)\Big)\;}\;}\,.\label{R-natural}
\eea
The matrix $\hPhi(i,t)$ is in this case conveniently defined in the mixed-basis representation, 
where $n$ in $\Phi_{n\bar{m}}(i,t)$ refers to a Fock state in the original basis, and $\bar{m}$ to a Fock state in the natural one.  Such a mixed-basis representation is useful since,throughout all calculations, one does not actually need to know what the natural basis is in terms of the original one\cite{NicolaPRB2008}. Such a nice property is linked to the gauge-invariance, equations \eqn{W-P} and \eqn{W-Psi}, of the theory\cite{RISB}. 

\subsection{The model}
We shall assume the generic Hamiltonian 
\be
\mathcal{H} = \sum_{i\not= j}\,\bo{c}^\dagger_{i}\, \hat{t}_{ij}\, \bo{c}^\dagger_j 
 \,+\, \sum_i\,\mathcal{H}_i
\;,\label{Ham}
\ee
where $\mathcal{H}_i$ includes all on-site terms. If the constraints Eq.~\eqn{1} and Eq.~\eqn{2} are satisfied at any time $t$, then, in infinite coordination lattices, it holds that\cite{BW&G1998,Fabrizio_review} 
\bea
E(t) &=& \bra{\Psi(t)}\mathcal{H}\ket{\Psi(t)}  = \bra{\Psi_0(t)}\mathcal{H}_*(t)\ket{\Psi_0(t)} \nonumber\\
&& + \sum_i\,\Tr\Big(\hPhi(i,t)^\dagger\,\hat{H}_i\,\hPhi(i,t)\Big)\nonumber\\
&& \equiv E_*(t) + \sum_i\,\Tr\Big(\hPhi(i,t)^\dagger\,\hat{H}_i\,\hPhi(i,t)\Big)\,,
\label{E}
\eea
where 
\bea
\mathcal{H}_*(t) &=&  \sum_{i\not=j}\,\mathbf{c}_i^\dagger\;\hat{R}(i,t)^\dagger\,
\hat{t}_{ij}\,\hat{R}(j,t)\;\mathbf{c}_i^\dagga
\;,\label{Ham*}
\eea
may be interpreted as the Hamiltonian of the quasiparticles. Evidently, all expectation values can be straightforwardly evaluated since the uncorrelated wavefunction $|\Psi_0(t)\rangle$ allows using Wick's theorem. 

\subsection{The action}

In the time-domain the variational principle corresponds to searching for the saddle point of the 
action\cite{SchiroFabrizioPRL} 
\bea
\mathcal{S} &=& \int dt \bigg[ i\,\bra{\Psi(t)} \dot{\Psi}(t)\rangle - E(t)\bigg]\nonumber\\
&\equiv& \int dt\, \Bigg\{ i\,\sum_i\, \Tr\left(\hat{\Phi}(i,t)^\dagger\,\fract{\partial \hat{\Phi}(i,t)}
{\partial t}\right) \nonumber\\
&&\qquad  \qquad \,+\, 
i\,\bra{\Psi_0(t)} \dot{\Psi}_0(t)\rangle 
\;-\; E(t)\Bigg\}\,,\label{S}
\eea
where the equivalence holds on provision that the constraints \eqn{1} and \eqn{2} are fulfilled at any time. 
The saddle point equations are readily obtained:
\bea
i\,\fract{\partial \hPhi(i,t)}{\partial t} &=& \hat{H}_i\,\hat{\Phi}(i,t) \,+\, \fract{\partial E_*(t)}{\partial \hat{\Phi}(i,t)^\dagger}\,,
\label{dot-Phi}\\
i\!\ket{\dot{\Psi}_0(t)} &=& \mathcal{H}_*(t)\ket{\Psi_0(t)}\,,\label{dot-Psi}
\eea
where 
\bea
\fract{\partial E_*(t)}{\partial \hat{\Phi}(i,t)^\dagger} &=& 
\Big\langle \Psi_0(t)\bigg| \fract{\partial \mathcal{H}_*(t)}{\partial \hat{\Phi}(i,t)^\dagger}\bigg|\Psi_0(t)\Big\rangle\nonumber\\
&& \equiv\hat{T}(i,t)\,\hPhi(i,t)\,.\label{T_*}
\eea
$\hat{T}(i,t)$ is a tensor with components $T_{nm;n'm'}(i,t)$, which is still functional of the 
matrices $\hPhi$ and $\hPhi^\dagger$ at site $i$ as well as at all sites connected to $i$ by the hopping. 
One can show that this tensor is hermitean, $\hat{T}(i,t)=\hat{T}(i,t)^\dagger$, which implies that the normalisation Eq.~\eqn{1} is conserved by the time evolution. 

\subsection{Fate of the constraint}

Concerning the second constraint, Eq.~\eqn{2}, we now prove that, if it is satisfied at the initial time, it will remain so at the saddle point solutions of Eq.~\eqn{dot-Phi} and Eq.~\eqn{dot-Psi}. 
Suppose we have indeed found the saddle point $\hat{\Phi}(i,t)$ and $|\Psi_0(t)\rangle$. By definition, any small variation with respect to that solution must lead to a vanishing variation of the action. Let us consider the infinitesimal gauge transformation 
\ba
\hat{\Phi}(i,t) + \delta\hat{\Phi}(i,t) &=& \hat{\Phi}(i,t)\,\Big(1-i\,\hat{K}(i,t)\Big),\\
\ket{\Psi_0(t)} + \ket{\delta\Psi_0(t)} &=& 
\left(1+i\sum_i\,\mathcal{K}(i,t)\right)\!\ket{\Psi_0(t)}\,,
\ea
where the operator 
\be
\mathcal{K}(i,t) = \sum_{\alpha\beta}\,K_{\alpha\beta}(i,t)\,c^\dagger_{i\,\alpha}\,c^\dagga_{i\,\beta}\,,\label{K}
\ee
has infinitesimal matrix elements $K_{\alpha\beta}(i,t)=K_{\beta\alpha}(i,t)^*$, and $\hat{K}(i,t)$ is its matrix representation in the Fock space. We already mentioned  that the energy $E(t)$ is gauge invariant 
so that the variation of the action, $\delta\mathcal{S} = \mathcal{S}^\cW - \mathcal{S}$, simply reads
\bw
\ba
\delta \mathcal{S}&=& 
\int dt\, \Bigg\{  i\,\sum_i\, \Tr\bigg(\hat{\Phi}(i,t)^\dagger\,\hat{\Phi}(i,t)^\dagga\, 
\fract{\partial \hat{W}(i,t)^\dagger}{\partial t}\,\hat{W}(i,t)^\dagga\bigg)
+ i\,\sum_i\, \bra{\Psi_0(t)} \mathcal{W}(i,t)^\dagger \, \dot{\mathcal{W}}(i,t)
\ket{\Psi_0(t)}\Bigg\}\\
&\simeq& \sum_i\, 
\int dt\, \Bigg\{  \Tr\bigg(\hat{\Phi}(i,t)^\dagger\,\hat{\Phi}(i,t)^\dagga\, \dot{K}(i,t)\bigg)\,-\, 
\bra{\Psi_0(t)} \dot{\mathcal{K}}(i,t)
\ket{\Psi_0(t)}\Bigg\}\\
&=&  \sum_i\, \sum_{\alpha\beta}\, \int dt\, \dot{K}_{\alpha\beta}(i,t)\,
\Bigg\{ \Tr\bigg(\hat{\Phi}(i,t)^\dagger\,\hat{\Phi}(i,t)^\dagga\, \hat{c}_{i\,\alpha}^\dagger \,
\hat{c}_{i\,\beta}^\dagger\bigg) - \bra{\Psi_0(t)} c_{i\,\alpha}^\dagger\,c_{i\,\beta}^\dagga
\ket{\Psi_0(t)}\Bigg\}\\
&=&  -\sum_i\, \sum_{\alpha\beta}\, \int dt\, K_{\alpha\beta}(i,t)\;\fract{\partial}{\partial t}
\Bigg\{ \Tr\bigg(\hat{\Phi}(i,t)^\dagger\,\hat{\Phi}(i,t)^\dagga\, \hat{c}_{i\,\alpha}^\dagger \,
\hat{c}_{i\,\beta}^\dagger\bigg) - \bra{\Psi_0(t)} c_{i\,\alpha}^\dagger\,c_{i\,\beta}^\dagga
\ket{\Psi_0(t)}\Bigg\}\,.
\ea
\ew
Since $\hat{\Phi}(i,t)$ and $|\Psi_0(t)\rangle$ are solutions of the saddle point equations, 
it follows that $\delta S$ 
must strictly vanish for any choice of the infinitesimally small matrix elements $K_{\alpha\beta}(t)$, which implies 
\ba
&&\fract{\partial}{\partial t}
\Bigg\{ \Tr\bigg(\hat{\Phi}(i,t)^\dagger\,\hat{\Phi}(i,t)^\dagga\, \hat{c}_{i\,\alpha}^\dagger \,
\hat{c}_{i\,\beta}^\dagger\bigg) \Bigg\}\\
&& \phantom{\fract{\partial}{\partial t}\Bigg\{}\qquad
- \bra{\Psi_0(t)} c_{i\,\alpha}^\dagger\,c_{i\,\beta}^\dagga
\ket{\Psi_0(t)}\Bigg\} =0\,,
\ea
thus just the desired result. It actually means that the term in parenthesis is conserved in the evolution. Therefore, if it is initially vanishing, it will remain so at any time, which thus implies that the constraint Eq.~\eqn{2} is fulfilled during the whole time evolution. \\


\subsection{Stationary problem}

At equilibrium one needs to find the minimum of the energy with the two constraints Eqs.~\eqn{1} and \eqn{2}, which can be enforced e.g. by Lagrange multipliers, leading to the set of equations 
\bea
\Lambda(i)\,\hat{\Phi}(i) &=& \Big(\hat{H}_i + \hat{T}(i)\Big)\,\hat{\Phi}(i) 
\nonumber\\
&& + \sum_{\alpha\beta}\,\mu_{\alpha\beta}(i)\,
\hat{\Phi}(i)\;\hat{d}_{i\alpha}^\dagger\,\hat{d}_{i\beta}^\dagga,\label{E-Phi}\\
E_* \ket{\Psi_0} &=& 
\Big(\mathcal{H}_* - \sum_i\, \mu_{\alpha\beta}(i)\;
d^\dagger_{i\,\alpha}\,d^\dagga_{i\,\beta}\Big)\ket{\Psi_0},
\label{E-Psi}
\eea
where $\Lambda(i)$ enforces Eq.~\eqn{1}, and the hermitean matrix $\hat{\mu}(i)$ 
with components $\mu_{\alpha\beta}(i)$ enforces Eq.~\eqn{2}. 
In whatever follows we shall assume to work in a mixed-basis representation where the operators $d^\dagga_{i\alpha}$ are associated to the natural orbitals, so that we must also ensure that 
\[
\Tr\Big(\hPhi^\dagger(i)\,\hat{\Phi}(i)\;\hat{d}_{i\alpha}^\dagger\,\hat{d}_{i\beta}^\dagga\Big) = 
\langle\Psi_0\mid d^\dagger_{i\alpha}d^\dagga_{i\beta}\mid\Psi_0\rangle = \delta_{\alpha\beta}\,n_\alpha(i)\,.
\]
The quasiparticle Hamiltonian in the natural basis, including explicitly the Lagrange multipliers,  is therefore 
\be
\mathcal{H}_* \to \sum_{i\not=j}\,\mathbf{d}_i^\dagger\;\hat{R}(i)^\dagger\,
\hat{t}_{ij}\,\hat{R}(j)\;\mathbf{d}_i^\dagga - \sum_i\, \mathbf{d}^\dagger_i\,\hat{\mu}(i)\,
\mathbf{d}^\dagga_i
\;,\label{H*-natural}
\ee
with $\hat{R}$ defined in Eq.~\eqn{R-natural}. Working in the mixed-basis representation with the natural orbitals considerably simplifies all  calculations. \\  
Recalling that $\hat{T}(i)$ is still functional of $\hPhi$, Eq.~\eqn{E-Phi} looks like a stationary non-linear Schr{\oe}dinger equation~\cite{NicolaPRB2012,NicolaPRX}. One can for instance solve it as in any Hartree-Fock calculation. Namely, one can find the eigenstates and eigenvalues of Eq.~\eqn{E-Phi} assuming $\hat{T}(i)$ fixed, and impose that, when $\hat{T}(i)$ is calculated substituting the actual expression of the lowest energy solution $\hPhi_0(i)$,  the two values coincide. The Lagrange multiplier $\hat{\mu}$ is fixed by imposing Eq.~\eqn{2} 
and Eq.~\eqn{natural}. In this way one finally gets the self-consistent $\hat{T}(i)$, which we shall hereafter denote as 
\be
\hat{T}^{(0)}(i) \equiv \hat{T}\Big[\hPhi_0^\dagga,\hPhi_0^\dagger\Big]\,.\label{T-0}
\ee
Once the latter is known, as well as the value of $\hat{\mu}$, one can also solve \eqn{E-Phi} for all  eigenvectors, $\hPhi_n(i)$ and corresponding eigenvalues $E_n(i)$, with $E_0(i)=\Lambda(i)$. 
We shall denote $\mathcal{H}_*$, $\hat{R}$, 
$\hat{Q}$, $\hat{n}$ and $\hat{S}$ calculated with $\hPhi_0$ as $\mathcal{H}^{(0)}_*$, $\hat{R}^\z$, $\hat{Q}^\z$, $\hat{n}^\z$ and $\hat{S}^\z$, respectively, with the latter two matrices diagonal in the natural basis,
\bea
n^\z_{\alpha\beta} &=& \delta_{\alpha\beta}\,n^\z_\alpha\,,
\label{n-diagonal}\\
S^\z_{\alpha\beta} &=& \delta_{\alpha\beta}\,S^\z_\alpha = 
\delta_{\alpha\beta}\,
\left(n^\z_\alpha\left(1-n^\z_\alpha\right)\right)^{-1/2}\,.
\label{S-diagonal}
\eea\\

We conclude by noting that the saddle point Hamiltonian Eq.~\eqn{H*-natural} with the inclusion of the Lagrange multipliers is not anymore invariant under the most general $U(2M)$ gauge transformation, but only under a subgroup $G$ with generators $\hat{T}^a$ that commute with $\hat{\mu}$. This is common in theories where the gauge invariance implements constraints about physical states. In the natural basis representation, 
$\mu_{i,\alpha\beta}=\delta_{\alpha\beta}\,\mu_{i\alpha}$ 
is diagonal, so that  the matrix elements of $\hat{T}^a$ must satisfy 
\be
T^a_{i,\alpha\beta}\,\big(\mu_{i\alpha}-\mu_{i\beta}\big)=0\,,
\ee
whose solution is straightforward. For any non-degenerate $\alpha$, i.e. such that 
$\mu_{i\alpha}\not = \mu_{i\beta}$, $\forall\,\beta\not=\alpha$, 
we associate the generators 
$T^\alpha_{i,\gamma\beta} = \delta_{\alpha\beta}\,\delta_{\gamma\beta}$ of $U(1)$ 
abelian groups. On the contrary, for any set of $\alpha_i$, 
$i=1,\dots,k$, such that $\mu_{i\alpha_i} = \mu_{i\alpha_j} 
\not = \mu_{i\beta}$, $\forall\, \beta\not= \alpha_1,\dots,\alpha_k$, 
we can associate generators of a $U(k)$ Lie algebra. \\

\section{Fluctuations above the saddle point solution}
\label{Fluctuations above the saddle point solution}

Our goal is to determine the action of the fluctuations beyond the saddle point within the harmonic approximation. To that purpose we assume that 
\be
\hPhi(i,t) = \text{e}^{-iE_0 t}\;\sum_n\,\phi_n(i,t)\,\hPhi_n(i)\;\hat{W}(i,t)^\dagger\,,\label{delta-Phi}
\ee
where $\phi_n(i,t)$ for $n>0$ is regarded as a first order fluctuation, while, to enforce normalisation, 
\be
\phi_0(i,t) = 1 - \fract{1}{2}\,\sum_{n>0}\,\left|\phi_n(i,t)\right|^2\,.
\ee
In addition, the Slater determinant is defined through 
\be
\ket{\Psi_0(t)} \to  \text{e}^{-iE_* t}\; \mathcal{W}(t)\,\ket{\Psi_0(t)}\,,
\label{delta-Psi}
\ee
where $|\Psi_0(t)\rangle$ is properly normalised and includes the zeroth order $|\Psi^\z_0\rangle$, solution of the saddle point, as well as 
a fluctuation correction $| \delta\Psi_0(t)\rangle$. The unitary operator 
\be
\mathcal{W}(i,t) = \exp\bigg(-i\,t\,\mathbf{d}^\dagger_i\,\hat{\mu}(i)\,\mathbf{d}_i\bigg)\,,\label{def:W}
\ee
where $\hat{\mu}(i)$ is the equilibrium Lagrange multiplier, and $\hat{W}(i,t)$ is the matrix representation
of $\mathcal{W}(i,t)$.\\
Through the above definitions, the action becomes
\bea
\mathcal{S}\! &=&\! \int \!dt \Bigg\{ i\sum_i\,\sum_{n>0}\, \phi_n(i,t)^*\,\dot{\phi}_n(i,t) 
+ i\,\langle\Psi_0(t)\mid \dot{\Psi}_0(t)\rangle\nonumber \\
&& \qquad \qquad - \sum_i\,\sum_{nm}\phi_n(i,t)^*\,V_{nm}(i)\,\phi_m(i,t)\nonumber\\
&& \qquad \qquad \qquad \qquad\;+E_0 + E_* -E_*(t)\;
\Bigg\},\;\;\;\;
\label{new-action}
\eea
where 
$
E_*(t) = \langle \Psi_0(t)\mid \mathcal{H}_*(t)\mid\Psi_0(t)\rangle
$,
being now 
\bea
\mathcal{H}_*(t) &=&  \sum_{i\not=j}\,\mathbf{d}_i^\dagger\;\hat{R}(i,t)^\dagger\,
\hat{t}_{ij}\,\hat{R}(j,t)\;\mathbf{d}_i^\dagga \!-\! \sum_i\,\bo{d}^\dagger_i\,\hat{\mu}(i)\,\bo{d}^\dagga_i,\qquad\label{new-H*}
\eea
and 
\bea
V_{nm}(i) &=& \Tr\Big(\hPhi_n(i)^\dagger\,\hat{H}_i\,\hPhi_m(i)\Big)
\label{V-nm(i)}\\
&& \qquad\qquad \qquad +  \Tr\Big(\hPhi_n(i)^\dagger\,\hPhi_m(i)\,\hat{\bo{d}}^\dagger_i\,\hat{\mu}(i)\,
\hat{\bo{d}}^\dagga_i\Big)\,.\nonumber
\eea

We expand $\mathcal{H}_*(t)$ up to second order in the fluctuations. 
The zeroth order is just $\mathcal{H}_*^\z$. 
Since the stationary solution is the saddle point of the action, the expectation value of the first order expansion 
$\mathcal{H}^{(1)}(t)$ over the saddle point Slater determinant 
$|\Psi^\z_0\rangle$ cancels with the first order expansion of the local energy $\sum_i\,\sum_{nm}\phi_n(i,t)^*\,V_{nm}(i)\,\phi_m(i,t)$. Therefore $\mathcal{H}^{(1)}(t)$ contributes to $E_*(t)$ with a second order term that, by linear response theory, reads 
\bea
\delta_1 E_*(t) &=& 
\langle \delta\Psi_0(t)\mid \mathcal{H}^{(1)}(t)\mid\Psi^\z_0\rangle 
+ c.c.\nonumber\\
&=& -i\,\int^t d\tau\, 
\big\langle\,\Big[\mathcal{H}^{(1)}(t)\,,\,
\mathcal{H}^{(1)}(\tau)\Big]\;\big\rangle_0\,,\label{delta E-1}
\eea
where, hereafter, $\langle\dots\rangle_0$ will denote average over 
$|\Psi^\z_0\rangle$, and the operators in Eq.~\eqn{delta E-1} have an additional time dependence since are evolved with the saddle point  
Hamiltonian $\mathcal{H}_*^\z$. The explicit expression of $\mathcal{H}^{(1)}(t)$ is
\bea
\mathcal{H}^{(1)}(t) &=& \sum_{i\not=j}\,\Bigg[
\bo{d}_j^\dagger\,\hat{R}^\z(j){^\dagger}\,
\hat{t}_{ji}\,\hat{R}^{(1)}(i,t)\,\bo{d}_i^\dagga + H.c.\Bigg],
\;\;\;\;\label{H-1}
\eea
where $\hat{R}^\z(i)$ is the stationary value, while  the explicit expression of the first order Taylor expansion $\hat{R}^{(1)}(i,t)$ is 
given in Appendix \ref{app-1-2}, see Eq.~\eqn{app-R-1-final}. \\

There are several second order terms upon expanding $\mathcal{H}_*(t)$, 
which we shall consider separately. The first is simply 
\bea
\mathcal{H}_1^{(2)}(t) &=& \sum_{i\not=j}\,
\bo{d}_i^\dagger\,\hat{R}^{(1)}(i,t){^\dagger}\,
\hat{t}_{ij}\,\hat{R}^{(1)}(j,t)\,\bo{d}_j^\dagga\,,\label{H2-1}
\eea
whose expectation value over $|\Psi^\z_0\rangle$ is an additional second order contribution 
\be
\delta_2 E_*(t) = \big\langle\;\mathcal{H}_1^{(2)}(t)\;\big\rangle_0
\,,\label{delta E-2}
\ee
which, together with $\delta_1 E_*(t)$ in Eq.~\eqn{delta E-1}, endow  the action with spatial correlations among the $\phi_n(i,t)$'s at different sites.\\
The next second order corrections to $\mathcal{H}_*(t)$ derive from the second order expansion of $\hat{R}(i,t)$
\bea
\hat{R}^{(2)}(i,t) &=& \hat{R}_1^{(2)}(i,t) + \hat{R}_2^{(2)}(i,t)\,,
\label{R-2-12}
\eea
where we distinguish two different contributions, see equations \eqn{app-R-2-1} and 
\eqn{app-R-2-2} in Appendix \ref{app-1-2}. The reason of this distinction is that
\bea
\delta_3 E_*(t) &=& \sum_i\,\sum_{nm}\phi_n(i,t)^*\,V_{nm}(i)\,\phi_m(i,t) \nonumber\\
&&\!\!+ \,\langle\;
\sum_{i\not=j}\, \Big(\bo{d}^\dagger_j\,\hat{R}^\z(j)\,\hat{t}_{ji}\,
\hat{R}_1^{(2)}(i,t)\,\bo{d}^\dagga_i + H.c.\Big)\;\rangle_0
\nonumber\\
&=& \sum_{n>0}\,\Big(E_n-E_0\Big)\,\phi_n(i,t)^*\,\phi_n(i,t)\,,
\label{delta E-3}
\eea
reproduces the bare excitation energy of the fluctuations. The last contribution to the energy of the fluctuations is therefore
\bea
\delta_4 E_*(t) &=& \!\langle\,\sum_{i\not=j} \Big(\bo{d}^\dagger_j\hat{R}^\z(j)\hat{t}_{ji}\,
\hat{R}_2^{(2)}(i,t)\bo{d}^\dagga_i \!+\! H.c.\Big)\rangle_0
.\qquad
\label{delta E-4}
\eea \\
If we define new variables 
\bea
x_n(i,t) &=& \fract{1}{\sqrt{2}}\,\Big(\phi_n(i,t) + \phi_n(i,t)^*\Big)\,,
\label{x_n}\\
p_n(i,t) &=& -\fract{i}{\sqrt{2}}\,\Big(\phi_n(i,t) - \phi_n(i,t)^*\Big)\,,
\label{p_n}
\eea
and the quadratic potential 
\be
U\big(t,\{x,p\}\big) = \delta_1 E_*(t) + \delta_2 E_*(t) + \delta_4 E_*(t)\,,
\label{V(t)-action}
\ee
which has a retarded component $\delta_1 E_*(t)$, see 
Eq.~\eqn{delta E-1}, the action of the fluctuations reads, upon defining $\omega_n = E_n-E_0$, 
\bea
\delta\mathcal{S} &=& \int dt\,\Bigg\{ 
\sum_i\,\sum_{n>0}\, \bigg[ p_n(i,t)\,\dot{x}_n(i,t) 
\label{action-fluctuations}\\
&& \!\!\!\!- \fract{\omega_n}{2}\,\Big(x_n(i,t)^2 + p_n(i,t)^2\Big)\bigg]
- U\big(t,\{x,p\}\big)\Bigg\},\nonumber
\eea  
which is just the action of coupled harmonic oscillators.\\
$\delta S$ in Eq.~\eqn{action-fluctuations} can be for instance used to evaluate the fluctuation corrections to linear response functions of local operators. For any local observable $\hat{O}(i)$, let us define the matrix element 
\be
O_n(i) \equiv \Tr\Big(\hPhi_n(i)^\dagger\,\hat{O}(i)\,\hPhi_0(i)\Big)\,.\label{def-On}
\ee
Suppose we add a perturbation that couples to the local density 
matrix 
\be
\delta\mathcal{H}(t) = 
\sum_i\,\bo{c}_i^\dagger\,\hat{V}(i,t)\,\bo{c}_i^\dagga\,,
\label{perturbation-V}
\ee
where the matrix $\hat{V}(i,t)$ with elements $V_{\alpha\beta}(i,t)$ represents the external field. Without loss of generality we can assume that the expectation value of $\delta\mathcal{H}(t)$ in Eq.~\eqn{perturbation-V} vanishes at the stationary solution. Since by assumption the external field is first order, the perturbation adds a second order correction to the action \eqn{action-fluctuations} that is 
\bea
V(t) &=& \sum_i\,\sum_n\,\Bigg[
\phi_n(i,t)^*\,\Tr\Big(\hPhi_n(i)^\dagger\,
\hat{\bo{c}}_i^\dagger\,\hat{V}(i,t)\,\hat{\bo{c}}_i^\dagga\,
\hPhi_0(i)\Big)\nonumber\\
&& + \phi_n(i,t)\,\Tr\Big(\hPhi_0(i)^\dagger\,
\hat{\bo{c}}_i^\dagger\,\hat{V}(i,t)\,\hat{\bo{c}}_i^\dagga\,
\hPhi_n(i)\Big)\Bigg]\label{perturbation-V-1}\\
&\equiv& \sum_i\,\sum_n\,\Big(\phi_n(i,t)^*\,V_n(i,t) 
+ \phi_n(i,t)\,V_n(i,t)^*\Big) \nonumber\\
&=& \sqrt{2}\sum_{i\,n}\bigg(\!
\Rea V_n(i,t)\,x_n(i,t) + \Ima V_n(i,t)\,p_n(i,t)\bigg).\nonumber
\eea
In the presence of $V(t)$ the action transforms into that of forced harmonic oscillators, whose solution allows calculating the expectation value of any local operator $\hat{O}(i)$, see Eq.~\eqn{def-On}, 
\begin{align*}
O(i,t) &= \Tr\Big(\hPhi(i,t)^\dagger\,\hat{O}(i)\,\hPhi(i,t)\Big)\\
& \simeq \sqrt{2}\sum_{n}\bigg(\!
\Rea O_n(i)\,x_n(i,t) + \Ima O_n(i)\,p_n(i,t)\bigg),
\end{align*}
at linear order in the external field. 

\subsection{Residual gauge invariance and would-be Goldstone modes}
\label{gauge fixing}
As we mentioned, the action Eq.~\eqn{new-action}, with the time dependent quasiparticle Hamiltonian defined in Eq.~\eqn{new-H*}, is invariant under a subgroup $G$ of the initial $U(2M)$ gauge symmetry. This implies the existence of massless modes with singular propagators that diverge as $1/\omega^2$ at low frequency, which are the would-be Goldstone modes related to the fact that the saddle-point $\hPhi_0(i)$ is not invariant under $G$. Let us consider for instance a $U(1)$ subgroup of $G$ related to the non-degenerate state $\alpha$ in the natural basis. The associated adjoint charge is 
\[
n_\alpha(i,t) \simeq \!\sum_{n>0} \left(\phi_n(i,t)^*\,\Tr\Big(\hPhi_n(i)^\dagger\,\hPhi_0(i)\,
\hat{d}^\dagger_{i\alpha}\,\hat{d}^\dagga_{i\alpha}\Big) + c.c.\right),
\]
and its conjugate variable is readily found to be
\begin{align*}
\varphi_\alpha(i,t) \simeq \fract{i}{2n^\z_\alpha}&\sum_{n>0} \bigg(\phi_n(i,t)^*\,\Tr\Big(\hPhi_n(i)^\dagger\,\hPhi_0(i)\,
\hat{d}^\dagger_{i\alpha}\,\hat{d}^\dagga_{i\alpha}\Big) \\
& 
- \phi_n(i,t)\,\Tr\Big(\hPhi_0(i)^\dagger\,\hPhi_n(i)\,
\hat{d}^\dagger_{i\alpha}\,\hat{d}^\dagga_{i\alpha}\Big)\bigg).
\end{align*}
The role of $\varphi_\alpha(i,t)$ is just to enforce the constraint 
Eq.~\eqn{2}, i.e. 
\[
n_\alpha(i,t) = \bra{\Psi_0(t)}c^\dagger_{i\alpha}c^\dagga_{i\alpha}
\ket{\Psi_0(t)}\equiv \langle\; c^\dagger_{i\alpha}c^\dagga_{i\alpha}\;\rangle_t\,.
\]
Indeed we can always perform a gauge transformation on the fermions
\[
c^\dagga_{i\alpha}\to \text{e}^{-i\varphi_\alpha(i,t)}\;
c^\dagga_{i\alpha}\,,
\]
which makes $\varphi_\alpha(i,t)$ to disappear from the energy leaving just the time derivative term in the action, 
\ba
\delta\mathcal{S}=-\int dt\,\dot{\varphi}_\alpha(i,t)\left(n_\alpha(i,t) -\langle\; c^\dagger_{i\alpha}c^\dagga_{i\alpha}\;\rangle_t\right)\,.
\ea
The condition of vanishing derivative with respect to $\varphi_\alpha(i,t)$ is therefore just the condition that the constraint is conserved.\\
It follows that we can always drop from the action all terms that contain the variables 
conjugate to the adjoint charges associated with the gauge symmetry $G$, on provision that, wherever $n_\alpha(i,t)$ appears, we replace   it with $\langle\; c^\dagger_{i\alpha}c^\dagga_{i\alpha}\;\rangle_t$.  
\\
However, the above procedure does not involve all the coefficients $\phi_n(i,t)$; some of their linear combinations are untouched by \textit{gauge-fixing} and remain genuine independent dynamical degrees of freedom\cite{Jolicoeur1991}. This fact, rather than being a limitation, it endows the theory with a richer dynamics.   

\section{Application to the half-filled Hubbard model}
\label{Application to the half-filled Hubbard model}

We now apply the above formalism to the simple case of a single band Hubbard model at half-filling, where all calculations can be worked out analytically and which also allows for a direct comparison with previous works\cite{VollhardtRMP,Jolicoeur1991,Raimondi1993,LorenzanaPRL,Lorenzana-PRB2003,LorenzanaPRB2004,SchiroFabrizioPRB,Sandri2012,Lorenzana&Capone2013}. We will show that we can indeed recover known results, but also find few novel ones.\\ 

The Hamiltonian is in this case 
\bea
\mathcal{H} &=& -\fract{t}{\sqrt{z}}\,\sum_{<ij>\sigma}\,
\Big(c^\dagger_{i\sigma} c^\dagga_{j\sigma} + H.c.\Big)\nonumber\\
&& + \fract{U}{4}\,\sum_i\,\bigg[2\Big(n_i-1\Big)^2 -1\bigg]\,,
\label{Ham-Hubbard}
\eea
where $<\!ij\!>$ means nearest neighbour bonds on a $d$-dimensional hyper cubic lattice, and $z=2d$ is the lattice coordination number that must be sent to $+\infty$ for the calculation to be really variational. \\
The local basis comprises four states which we choose to be, in order, the empty configuration, $\ket{\!0}$, the doubly occupied one, 
$\ket{\!2}$, the singly occupied by a spin up electron, $\ket{\,\up}$, and that occupied by a spin down one, $\ket{\,\down}$. The most general charge-conserving $\hPhi$ has the following form, dropping for the meanwhile the site index, 
\be
\hPhi = \fract{1}{\sqrt{2}}
\begin{pmatrix}
\hPhi_c & 0\\
0 &\hPhi_s
\end{pmatrix}\,,\label{Phi-Hubbard}
\ee 
where the charge component, i.e. the matrix elements in the subspace 
$\big(\ket{0},\ket{2}\big)$, is
\be
\hPhi_c = \begin{pmatrix}
\phi_{c0} + \phi_{c3} & 0\\
0 & \phi_{c0} - \phi_{c3}
\end{pmatrix} = \phi_{c0}\,\sigma_0 + \phi_{c3}\,\sigma_3\,,
\label{Phi-c}
\ee
with $\sigma_0$ the $2\times 2$ identity matrix, and $\sigma_i$, 
$i=1,\dots,3$ the Pauli matrices, whereas the spin component, namely 
the matrix elements in the subspace 
$\big(\ket{\,\up},\ket{\,\down}\big)$, is instead 
\be
\hPhi_s = \sum_{i=0}^3\,\phi_{si}\,\sigma_i 
= \phi_{s0}\,\sigma_0 + \bphi_s\cdot\bsigma\,,\label{Phi-s}
\ee
which allows a full spin-$SU(2)$ invariant 
analysis\cite{Wolfle1989,LorenzanaPRB2004}. Normalisation implies that 
\[
1 = \big|\phi_{c0}\big|^2 + \big|\phi_{c3}\big|^2 + \big|\phi_{s0}\big|^2 + \bphi_s^*\cdot\bphi_s\,.
\]
One can readily verify that the matrix $\hat{Q}$ with components
\be
Q_{\sigma\sigma'} = \Tr\Big(\hPhi^\dagger\,c^\dagga_{\sigma}\,\hPhi\,c^\dagger_{\sigma'}\Big)\,,\label{Q-Hubbard}
\ee
can be written as 
\be
\hat{Q} = Q_0\,\sigma_0 + \mathbf{Q}\cdot\bsigma\,,\label{Q-Hubbard-1}
\ee
where 
\begin{align}
2Q_0 &= \Big(\phi_{c0}^*\,\phi_{s0}^\dagga + \phi_{s0}^*\,\phi_{c0}^\dagga\Big)
+  \Big(\phi_{c3}^*\,\phi_{s0}^\dagga - \phi_{s0}^*\,\phi_{c3}^\dagga\Big)\,,\label{Q-0}\\
2Q_i &= \Big(\phi_{c0}^*\,\phi_{si}^\dagga - \phi_{si}^*\,\phi_{c0}^\dagga\Big)
+  \Big(\phi_{c3}^*\,\phi_{si}^\dagga + \phi_{si}^*\,\phi_{c3}^\dagga\Big)\,,\label{Q-i}
\end{align}
with $i=1,\dots,3$. 
Seemingly, 
\bea
\hat{\Delta} &\equiv& \Big(\phi_{c0}^*\,\phi_{c3}^\dagga + \phi_{c3}^*\,\phi_{c0}^\dagga\Big)\,\sigma_0 \nonumber\\
&& - \Big(\phi_{s0}^*\,\bphi_s + \phi_{s0}^\dagga\,\bphi_s^* + 
i\,\bphi_s^*\wedge\bphi_s^\dagga\Big)\cdot\bsigma\nonumber\\
&& \equiv 
\Delta_0\,\sigma_0 + \boldsymbol{\Delta}\cdot\bsigma\,.
\label{Delta}
\eea

\subsection{Stationary solution}
As common when discussing the Mott transition in the single band Hubbard model, we shall be interested in the stationary solution within the paramagnetic sector, i.e. neglecting spontaneous breakdown of spin $SU(2)$ symmetry.
Such solution at half-filling is characterised by a site independent 
\ba
\hPhi_0 = \fract{1}{\sqrt{2}}\;\begin{pmatrix}
\phi_{c0}^\z\,\sigma_0 & 0\\
0 & \phi_{s0}^\z\,\sigma_0
\end{pmatrix}\,,
\ea
with 
\[
1 = \big|\phi_{c0}^\z\big|^2 + \big|\phi_{s0}^\z\big|^2\,.
\]
Under this assumption 
\be
\hat{R}(i) = \Big(\phi_{c0}^\z{^*}\,\phi_{s0}^\z + 
\phi_{s0}^\z{^*}\,\phi_{c0}^\z\Big)\,\sigma_0 = R^\z\,\sigma_0\,,\;\;\forall\,i\,,
\label{R0-1}
\ee
so that the quasiparticle Hamiltonian is just a tight-binding model with renormalised hopping, i.e. 
\be
\mathcal{H}_*^\z = -\fract{t}{\sqrt{z}}\;R^\z{^2}\,
\sum_{<ij>}\,\Big(\bo{c}_i^\dagger\,\bo{c}_j^\dagga + H.c.\Big)\,,
\label{H*0-Hubbard}
\ee
and natural and original orbitals coincide.  
It follows that the stationary Slater determinant is the non-interacting Fermi sea. We define 
\ba
- \sum_i\, T_0 &\equiv&-\fract{t}{\sqrt{z}}\, \sum_{<ij>\, \sigma}\,\langle\; c^\dagger_{i\sigma}\,c^\dagga_{j\sigma}+H.c.\;\rangle_0  \,,
\ea
where $\langle\dots\rangle_0$ is the average over the Fermi sea. Therefore $-T_0$ is the hopping energy per site, and $-2T_0/z$ the hopping energy per bond of the Fermi sea. \\
The saddle point equations for $\hPhi_0$ can be readily found
\ba
E\,\phi_{c0}^\z &=& - 2T_0\,R^\z\,\phi_{s0} + \fract{U}{4}\;\phi_{c0}\,\\
E\,\phi_{s0}^\z &=& - 2T_0\,R^\z\,\phi_{c0} - \fract{U}{4}\;\phi_{s0}\,.
\ea
The lowest energy eigenvalue is  
\be
E_0 = -\fract{1}{2}\sqrt{U^2 + \Big(8T_0\,R^\z\Big)^2\;}\;,\label{E0-Hubbard}
\ee
and is characterised by 
\[
\phi_{c0}^\z = \sin\fract{\theta}{2}\,,\qquad
\phi_{s0}^\z = \cos\fract{\theta}{2}\,,
\]
with $\tan\theta = 8T_0\,R^\z/U$. 
Since through Eq.~\eqn{R0-1} $R^\z = \sin\theta$, the self-consistency condition implies 
\be
\tan\theta = \fract{8T_0\,R^\z}{U}= \fract{8T_0}{U}\;\sin\theta\,,\label{self-consistency}
\ee
namely
\be
\cos\theta = 
\begin{cases}
U/U_c & U\leq U_c = 8T_0\,,\\
1 & U > U_c\,.
\end{cases}\label{theta-Hubbard}
\ee
$U_c$ is the well known value of the Brinkman-Rice\cite{Brinkman&Rice} metal-insulator transition 
within the Gutzwiller approximation.\\ 
In conclusion, the lowest energy eigenstate is 
\bea
\hPhi_0 = \fract{1}{\sqrt{2}}\;\begin{pmatrix}
\sin\fract{\theta}{2}\,\sigma_0 & 0\\
0 & \cos\fract{\theta}{2}\,\sigma_0
\end{pmatrix}\,,\label{Phi0-Hubbard}
\eea  
where $\cos\theta = \text{min}\left(1,U/U_c\right)$, 
and has eigenvalue 
\be
E_0 = -\fract{U}{4\cos\theta} = -\fract{\text{Max}\left(U,U_c\right)}{4}\;.
\label{E0-Hubbard-1}
\ee\\
We can now find all other eigenvalues and eigenvectors. The highest energy one is 
\be
\hPhi_3 = \fract{1}{\sqrt{2}}\;\begin{pmatrix}
\cos\fract{\theta}{2}\,\sigma_0 & 0\\
0 & -\sin\fract{\theta}{2}\,\sigma_0
\end{pmatrix}\,,\label{Phi3-Hubbard}
\ee  
with eigenvalue 
\be
E_3 = -E_0\,.\label{E3-Hubbard}
\ee
This eigenstate actually corresponds to the high energy Hubbard bands.\\
The lowest excited eigenstate is threefold degenerate ($i=1,2,3$)
\be
\hPhi_{1\,i} = \fract{1}{\sqrt{2}}\;\begin{pmatrix}
0 & 0\\
0 & \sigma_i
\end{pmatrix}\,,\label{Phi1-Hubbard}
\ee  
with eigenvalue 
\be
E_1 = -\fract{U}{4}\,,\label{E1-Hubbard}
\ee
and describes spin fluctuations. We note that above the Brinkmann-Rice transition, $U>U_c$, this magnetic state becomes degenerate with the ground state. In what follows we shall anyway expand always around $\hPhi_0$, and, to avoid problems, we will mostly consider the metal phase at $U\leq U_c$.  \\
Finally, the last eigenstate is 
\be
\hPhi_{2} = \fract{1}{\sqrt{2}}\;\begin{pmatrix}
\sigma_3 & 0\\
0 & 0
\end{pmatrix}\,,\label{Phi2-Hubbard}
\ee  
with eigenvalue 
\be
E_2 = +\fract{U}{4}\,,\label{E2-Hubbard}
\ee
and describes instead charge fluctuations. This mode becomes degenerate with $\hPhi_3$ above the transition.\\

\subsection{Action of the fluctuations}

Following section \ref{Fluctuations above the saddle point solution} we write 
\bea
\hPhi(i,t) &=& \phi_0(i,t)\,\hPhi_0 + \sum_{i=1}^3\,\phi_{1i}(i,t)\,\hPhi_{1\,i} \nonumber\\
&& + \sum_{n=2}^3\,\phi_n(i,t)\,\hPhi_n \,,\label{new-Phi}
\eea
with $\phi_0(i,t)$ fixed by normalisation. 
Through equations \eqn{Q-Hubbard-1}, \eqn{Q-0} and \eqn{Q-i} we find that 
\bea
\hat{R}^{(1)}(i,t) &=& \sin\fract{\theta}{2}\,\Big(\bphi_{1}(i,t)-\bphi_{1}(i,t)^*\Big)\cdot\bsigma
\nonumber\\
&& -\cos\fract{\theta}{2}\,\Big(\phi_2(i,t)-\phi_2(i,t)^*\Big) \nonumber
\\
&& 
+ \cos\theta\,\Big(\phi_3(i,t) + \phi_3(i,t)^*\Big)\nonumber\\
&\equiv& i\,\sqrt{2}\;\sin\fract{\theta}{2}\,
\mathbf{p}_1(i,t)\cdot\bsigma -i\,\sqrt{2}\;\cos\fract{\theta}{2}\,p_2(i,t)
\nonumber\\
&& + \sqrt{2}\;\cos\theta\,x_3(i,t)\,,\label{delta R-1}
\eea
where we have introduced the conjugate variables associated with $\phi_n$ and 
$\phi_n^*$. Eq.~\eqn{H-1} reads explicitly 
\bea
\mathcal{H}^{(1)}_* &=& \sum_i\,\Bigg\{
\sqrt{2}\;\left(2\cos\fract{\theta}{2}\right)^{-1}\,\boldsymbol{\nabla}\cdot\mathbf{J}_s(i)\cdot\mathbf{p}_1(i,t)
\nonumber\\
&& \qquad 
-\sqrt{2}\;\left(2\sin\fract{\theta}{2}\right)^{-1}\,\boldsymbol{\nabla}\cdot\mathbf{J}_c(i)\,p_2(i,t)\nonumber \\
&& \qquad   \qquad  + 2\sqrt{2}\;\cot\theta\,h_*(i)\,
x_3(i,t)\;\Bigg\}\,,
\label{H(1)-Hubb}
\eea
where $\boldsymbol{\nabla}$ is the lattice divergence, 
$\mathbf{J}_s(i)$ and $\mathbf{J}_c(i)$ the spin and charge currents, respectively, defined through the continuity equations 
\bea
i\,\fract{\partial}{\partial t}\Big( \bo{c}^\dagger_i\,\sigma_0\,\bo{c}^\dagga_i\Big)
 &=& 
\Big[\bo{c}^\dagger_i\,\sigma_0\,\bo{c}^\dagga_i\,,\,\mathcal{H}_*^{(0)}\Big]
\equiv -i \boldsymbol{\nabla}\cdot\mathbf{J}_c(i),
\;\;\;\;\;\label{J-c}
\\
i\,\fract{\partial}{\partial t}\Big( \bo{c}^\dagger_i\,\bsigma\,\bo{c}^\dagga_i\Big)&=& \Big[\bo{c}^\dagger_i\,\bsigma\,\bo{c}^\dagga_i\,,\,\mathcal{H}_*^{(0)}\Big]
\equiv -i \boldsymbol{\nabla}\cdot\mathbf{J}_s(i)\,.\;
\;\;\;\;\label{J-s}
\eea 
and finally $h_*(i)$ the Hamiltonian density 
\be
h_*(i) = -\fract{t}{2\sqrt{z}}\;R^\z{^2}\sum_{j\text{~n.n.~} i}\,
\Big(\bo{c}^\dagger_i\,\bo{c}^\dagga_j + H.c.\Big)\,.
\ee
Therefore $\delta_1 E_*(t)$ defined in Eq.~\eqn{delta E-1} becomes, due to particle-hole and spin $SU(2)$ symmetry
\bw
\bea
\delta_1 E_*(t) &=& \sum_{i,j}\,\int d\tau\,\Bigg\{
\fract{1}{1+\cos\theta}\;\chi_{_{\boldsymbol{\nabla J \nabla J}}}(i-j,t-\tau)\,
\mathbf{p}_1(i,t)\cdot\mathbf{p}_1(j,\tau)
+ \fract{1}{1-\cos\theta}\;\chi_{_{\boldsymbol{\nabla J \nabla J}}}(i-j,t-\tau)\,
p_2(i,t)\,p_2(j,\tau)
\nonumber\\
&&\phantom{\sum_{i,j}\,\int d\tau\,\Bigg\{}\qquad
+ 8\,\cot^2\theta\,\chi_{h_* h_*}(i-j,t-\tau)\,x_3(i,t)\,x_3(j,\tau)\;\Bigg\}\,,
\label{delta E-1-Hub}
\eea
\ew
where $\chi_{_{\boldsymbol{\nabla J \nabla J}}}$ is the linear response function of $\boldsymbol{\nabla J}$ 
with the Hamiltonian $\mathcal{H}^\z_*$, which is actually the same for charge and spin currents, and $\chi_{h_* h_*}$ the response function of $h_*$. We observe that, because of charge and spin continuity equations, in Fourier space the following equivalence holds
\be
2T_0\,\sin^2\theta\,\Big(\gamma_\mathbf{0}-\gamma_\bq\Big)
+\chi_{_{\boldsymbol{\nabla J \nabla J}}}(\bq,\omega) 
= \omega^2\,\chi(\bq,\omega),\label{gauge-invariance-1}
\ee
where $\chi(\bq,\omega)$ is the density-density response function, which is the same both in the charge and spin channels, and by definition 
\be
\gamma_\bq = \fract{2}{z}\,\sum_{i=1}^d\,\cos q_i\,\in\,[-1,+1]\,.\label{gamma-def}
\ee\\
Without going into further details, we find that the following expressions for the remaining contributions 
$\delta_2 E_*(t)$ in Eq.~\eqn{delta E-2}, and $\delta_4 E_*(t)$ in 
Eq.~\eqn{delta E-4}:
\bw
\bea
\delta_2 E_*(t) &=& -\fract{4T_0}{z}\,\sum_{<ij>}\, \Bigg\{ \sin^2\fract{\theta}{2}\;\mathbf{p}_1(i,t)
\cdot\mathbf{p}_1(j,t)
+ \cos^2\fract{\theta}{2}\;p_2(i,t)\,p_2(j,t) + \cos^2\theta\,x_3(i,t)\,x_3(j,t)\Bigg\},\\
\label{delta E-2-Hub}
\delta_4 E_*(t) &=& -2T_0\,\sin^2\theta\,\sum_i\,\Bigg\{\cos^2\fract{\theta}{2}\,\mathbf{x}_1(i,t) \cdot\mathbf{x}_1(i,t) 
+  \sin^2\fract{\theta}{2}\,x_2(i,t) ^2
\Bigg\}.\label{delta E-4-Hub}
\eea
\ew
We have now all ingredients required to evaluate linear response functions of local operators within the harmonic approximation for the fluctuations. 

\subsection{Hubbard-band dispersion mode}
\label{sec-Hbands}
As we mentioned, the Hubbard bands may be associated with the excited state $\hPhi_3$, hence with  
the operators $x_3$ and $p_3$. Their equations of motion in Fourier space are 
\bea
-i\omega\,x_3(\bq,\omega) &=& \omega_3\,p_3(\bq,\omega)\,,\label{x3-dot}\\
-i\omega\,p_3(\bq,\omega) &=& -\Big[\,\omega_3
- 4T_0\,\gamma_\bq \nonumber\\
&& \phantom{-\Bigg(}
+ 8\cot^2\theta\,\chi_{h_* h_*}(\bq,\omega)\Big]\,x_3(\bq,\omega)
.\qquad\label{p3-dot}
\eea
Within the metal phase, $U<U_c$, $\omega_3=E_3-E_0 = 4T_0$, so that, upon defining $\cos\theta =U/U_c
\equiv u$, 
and noting that, for small $|\bq|$, $\chi_{hh}(\bq,\omega) = O(q^4)$, 
the eigenmode energy is solution of the equation 
\bea
\omega_{3\bq}^2 &=& 4T_0\,
\bigg[4T_0\big(1-u^2\big)
+4T_0\,u^2\,\big(\gamma_{\mathbf{0}}-\gamma_\bq\big) 
\nonumber\\
&& \qquad \qquad \qquad + 8\fract{u^2}{1-u^2}\, 
\chi_{h_*h_*}(\bq,\omega_{3\bq})
\bigg]
\label{x3-mode}
\\
&\simeq& 16T_0^2\Bigg[\big(1-u^2\big) + u^2\,
\big(\gamma_{\mathbf{0}}-\gamma_\bq\big)\Bigg] 
\,,
\nonumber
\eea
thus describes an optical mode that softens at the metal insulator transition, $\omega_{3\mathbf{0}} = 4T_0\,\sqrt{1-u^2\,}\,\to 0$ when $u\to 1$. We observe that the continuum of quasiparticle-quasihole excitations extends up to an energy of order $T_0\big(1-u^2\big)$, so that, upon approaching the transition, $\omega_{3\mathbf{q}}$ must detach from the continuum and become a genuine coherent excitation.\\
This coherent mode actually corresponds to the spin-wave excitations of the Ising field within the $Z_2$ slave-spin representation of the Hubbard model\cite{Z2-1,Z2-2,SchiroFabrizioPRB}. This is not surprising since, as shown in Ref.~\onlinecite{SchiroFabrizioPRB}, the Gutzwiller wavefunction is just the mean-field variational state of the $Z_2$ slave-spin theory. At the mean-field level, the Mott transition in this representation translates into the order-disorder transition of a quantum Ising model. Therefore the mode $x_3$ seems to be  the real fingerprint of the Mott transition. \\

\subsection{Dynamical charge susceptibility}
\label{sec-charge}

We assume to perturb the system in the metal phase, $u\leq 1$, by an external potential that couples to the charge deviation from half-filling, namely
\bea
\delta \mathcal{H}(t) &=& \sum_i\,v(i,t)\,\big(n_i-1\big)\nonumber\\
&\simeq& -\sqrt{2}\;\sin\fract{\theta}{2}\,\sum_i\,v(i,t)\,x_2(i,t)\,.
\label{charge-density}
\eea
Since $\omega_2=E_2-E_0=2T_0\big(1+u\big)$ and by means of 
Eq.~\eqn{gauge-invariance-1}, we find  
in the presence of the field the following equations of motion for the conjugate variables $x_2$ and $p_2$ 
\ba
-i\omega\,x_2(\bq,\omega) &=& \fract{\omega^2}{1-u}\; \chi(\bq,\omega)\,p_2(\bq,\omega)\,,\\
-i\omega\,p_2(\bq,\omega) &=& 
\sqrt{2}\;\sin\fract{\theta}{2}\,v(\bq,\omega)\\
&& -2T_0\,\big(1+u\big)\,u\,\big(2-u\big)\,x(\bq,\omega) 
\,,
\ea
from which it follows that the dynamical charge susceptibility is 
\bea
\chi_c(\bq,\omega) 
&=& \fract{(1-u)\,\chi(\bq,\omega)}{\;(1-u)-2T_0\,\big(1+u\big)\,u\,\big(2-u\big)\,\chi(\bq,\omega)\;}
\nonumber\\
&\equiv& \fract{\chi(\bq,\omega)}{1+\Gamma_c\,\chi(\bq,\omega)}
\,,\label{charge-susceptibility}
\eea
where it is evident the analogy with conventional RPA, though with a renormalised 
coupling constant 
\bea
\Gamma_c &=& - \fract{U}{2}\,\fract{1+u}{1-u}\,\left(1-\fract{u}{2}\right)<0\,.\label{Gamma_c}
\eea
We note that  
\[
\chi(\bq\to 0,\omega=0) = -\mathcal{N}_*\,,
\]
where 
\be
\mathcal{N}_* = \fract{\mathcal{N}_0}{1-u^2}\,,
\label{quasiparticle DOS}
\ee
is the quasiparticle density of states (DOS) at the chemical potential, as opposed to the bare DOS $\mathcal{N}_0$, and diverges approaching the Mott transition. Therefore, through Eq.~\eqn{charge-susceptibility}, the charge compressibility is readily obtained
\bea
\kappa &=&  \fract{\mathcal{N}_*}{\;
1-\Gamma_c\,\mathcal{N}_*\;} \equiv \fract{\mathcal{N}_*}{1+F^S_0}\,,
\label{compressibility}
\nonumber
\eea
and defines the Landau $F_0^S$ parameter 
\be
F^S_0 = -\mathcal{N}_*\,\Gamma_c
\;.\label{F_0^S}
\ee 
Since approaching the transition, $u\to 1$, $F^S_0 \sim (1-u)^{-2}$ diverges faster than 
$\mathcal{N}_*\sim (1-u)^{-1}$, we find that the charge compressibility correctly vanishes at the MIT. 
The expression of $F^S_0$ coincides with that originally obtained by Vollhardt\cite{VollhardtRMP}. 
\\
In the opposite limit of small $|\bq|$ with respect to frequency,   
\[
\chi(\bq,\omega) \simeq \fract{2T_0\,\big(1-u^2\big)
\big(\gamma_{\mathbf{0}} - \gamma_\bq\big)}{\omega^2}\,, 
\]
which, inserted into Eq.~\eqn{charge-susceptibility}, allows calculating the poles of the dynamical charge susceptibility, 
which are 
\be
\omega_{c\bq}^2 = 4T_0^2\,(1+u)^2\,u\,(2-u)\,\big(\gamma_\mathbf{0}-\gamma_\bq\big)\,.\label{zero sound}
\ee
This acoustic mode is above the quasiparticle-quasihole continuum and actually corresponds to the Landau's zero sound. Once again this result is compatible with Vollhardt's description of the correlated metal within the Gutzwiller approximation in the framework of 
Landau-Fermi liquid theory\cite{VollhardtRMP}. Indeed the zero sound velocity has the expected Landau's expression, once one realises that 
in a lattice with infinite coordination $F_1^S=0$ and it is unrelated to the enhancement of the effective mass. \\ 
We conclude highlighting that the velocity of the zero sound stays constant approaching the Mott transition. In particular, for $\omega^2 \gg T_0\,\big(1-u^2\big)\,\big(\gamma_\mathbf{0}-\gamma_\bq\big)$, the dynamical charge susceptibility can be written as 
\be
\chi_c(\bq\to 0, \omega) = \fract{2T_0\,(1-u^2)\,\big(\gamma_\mathbf{0}-\gamma_\bq\big)}
{\omega^2 - \omega_{c\bq}^2}\;,
\ee
hence the pole at the zero sound has vanishing weight as the transition $u\to 1$ is approached, in agreement with the expectation that spectral weight is transferred at high energy. \\
We conclude by observing that the propagator $\Pi_2(\bq,\omega)$ of $p_2(\bq,\omega)$ 
\[
\Pi_2(\bq,\omega) = -\fract{1}{\omega^2}\;\fract{(1-u)\Gamma_c}
{1+\Gamma_c\,\chi(\bq,\omega)}\;,
\]
is singular at $\omega=0$, although this singularity does not appear in the physical response function, which is proportional to the propagator of the conjugate variable $x_2(\bq,\omega)$. Indeed, 
$p_2(\bq,\omega)$ is one of the would-be Goldstone modes that we mentioned in section~\ref{gauge fixing}. The action of the single-band Hubbard model is $U(2)=U(1)\times SU(2)$ gauge invariant, and 
$p_2(\bq,\omega)$ is just the would-be Goldstone mode associated with the abelian $U(1)$, whereas we shall see that $\mathbf{p}_1(\bq,\omega)$ are instead those associated with $SU(2)$. In fact, the RPA form of the charge susceptibility could be very easily obtained by the gauge-fixing prescription of section~\ref{gauge fixing}. If we drop all terms that contain $p_2(i,t)$ and replace 
\[
-\sqrt{2}\;\sin\fract{\theta}{2}\;x_2(i,t) \to \langle\; n_i-1\;\rangle_t\,,
\]  
we get an effective Hamiltonian of the quasiparticles, neglecting for convenience all other variables but $x_2(i,t)$,  
\ba
\mathcal{H}_*(t) &=& \mathcal{H}^\z_* +\sum_i\,v_*(i,t)\,\big(n_i-1\big)\,,
\ea
where 
\be
v_*(i,t) = v(i,t) -\Gamma_c\,\langle\; n_i-1\;\rangle_t\,,
\label{eff-v}
\ee
which readily leads to Eq.~\eqn{charge-susceptibility}.

\subsection{Dynamical spin susceptibility}
\label{sec-spin}
In order to study the spin response, we imagine to add an external field that couples to the spin density, e.g. to its $z$ component, namely
\bea
\delta \mathcal{H}(t) &=& -\sum_i\,B_3(i,t)\,\big(n_{i\up}
- n_{i\down}\big)\nonumber\\
&=& -\sqrt{2}\;\cos\fract{\theta}{2}\,\sum_i\,B_3(i,t)\,x_{1,3}(i,t)\,.
\label{spin-density}
\eea
In the metal phase $\omega_1=E_1-E_0=2T_0\big(1-u\big)$, and repeating all calculations done for the charge susceptibility, we finally obtain the dynamical spin susceptibility
\bea
\chi_s(\bq,\omega) &=&  \fract{\chi(\bq,\omega)}{1 + \Gamma_s\,\chi(\bq,\omega)}\;,
\label{spin-susceptibility}
\eea
where 
\be
\Gamma_s = \fract{U}{2}\,\fract{1-u}{1+u}\,\left(1+\fract{u}{2}\right)>0\,.\label{Gamma_s}
\ee
The above expression reproduces the small $u$ Stoner's enhancement of the magnetic susceptibility. In addition it satisfies the relationship $\Gamma_s(U) = \Gamma_c(-U)$ valid at particle-hole symmetry\cite{VollhardtRMP}. 
Since $\Gamma_s\sim (1-u)$ vanishes linearly approaching the transition, the Landau's parameter 
\be
F^A_0 = - \mathcal{N}_*\,\Gamma_s<0\,,\label{F_0^A}
\ee
is constant for $u\to 1$, which implies that the uniform static spin susceptibility diverges at the MIT. This 
result agrees with previous ones\cite{VollhardtRMP,LorenzanaPRB2004} also obtained within the Gutzwiller approximation, but contrasts DMFT, which instead finds a finite uniform spin susceptibility at the transition. \\
Such negative outcome critically depends from the fact that the effective interaction $\Gamma_s$, 
Eq.~\eqn{Gamma_s}, vanishes at the transition. We are going to show that beyond the harmonic approximation this cancellation does not occur anymore.\\
We note that $p_{1a}(i,t)$, $a=1,\dots,3$, are now the Goldstone modes associated with $SU(2)$ gauge invariance, and their propagators
\[
\Pi_{1a}(\bq,\omega) = -\fract{1}{\omega^2}\;
\fract{(1+u)\,\Gamma_s}{1+\Gamma_s\,\chi(\bq,\omega)}\;,
\] 
diverge at $\omega=0$. We can, as in section~\ref{sec-charge}, drop $\mathbf{p}_{1}(i,t)$ from the action and replace 
\[
\sqrt{2}\,\cos\fract{\theta}{2}\,\mathbf{x}_1(i,t) 
\to \langle\;\mathbf{c}^\dagger_i\,\boldsymbol{\sigma}\,
\mathbf{c}^\dagga_i\;\rangle_t\,,
\] 
whose effect could be absorbed into an effective magnetic field 
\be
B_{*a}(i,t) = \delta_{a3}\,B_3(i,t) - \Gamma_s\, \langle\;\mathbf{c}^\dagger_i\,\sigma_a\,
\mathbf{c}^\dagga_i\;\rangle_t\,,\label{eff-B}
\ee
that straightforwardly leads to Eq.~\eqn{spin-susceptibility}.

\subsection{Beyond RPA in the $x_3$ mode}
We observe that all the above results in the metal phase correspond to expanding the action at second order in the fluctuations but treating the linear coupling between the latter and the fermions just within RPA,  i.e. not accounting for exchange processes. 
While this procedure is somehow forced by gauge invariance for what it concerns charge and spin modes, see the ending parts of sections 
\ref{sec-charge} and \ref{sec-spin}, it is not really compulsory for the $x_3(i,t)$ mode that describes the Hubbard bands. We can therefore take a first step forward when dealing with $x_3(i,t)$ in the direction of the so called RPA+Exchange. According to Eq.~\eqn{H(1)-Hubb}, promoting $x_3$ and $p_3$ to quantum conjugate variables, after defining $t_*=t\,\sin^2\theta$ and 
\[
X(i) = 1+\sqrt{2}\;\cot\theta\;x_3(i)\,,
\]
the Hamiltonian reads 
\begin{align}
\mathcal{H}_* =& -\fract{t_*}{\sqrt{z}}\sum_{<ij>} \,\Big(\bo{c}_i^\dagger \bo{c}_j^\dagga+H.c.\Big)X(i)\,X(j)\label{H-with 3}\\
& + \sum_i\, \bigg[\Big(v_*(i,t)\,\bo{c}^\dagger_i\,\sigma_0\,\bo{c}^\dagga_i
+ \mathbf{B}_*(i,t)\cdot
\bo{c}^\dagger_i\,\boldsymbol{\sigma}\,\bo{c}^\dagga_i\Big) \nonumber\\
& + \fract{\omega_3}{2}\,\left(x_3(i)^2+p_3(i)^2\right) 
+ T_0\,\sin2\theta\,\sqrt{2}\;x_{3}(i)\bigg]
,\nonumber
\end{align}
where the effective fields are those in Eqs.~\eqn{eff-v} and \eqn{eff-B}. The last term in Eq.~\eqn{H-with 3}, linear in $x_3$, derives from Eq.~\eqn{V-nm(i)} and cancels the linear term of the hopping when the latter is averaged over the Fermi sea, which is just the saddle point condition for $x_3$.  \\
Near the Mott transition from the metal side, $u\lesssim 1$, since $t_*$ is small with respect to  $\omega_3$, we can integrate out $x_3$ and neglect the frequency dependence of its propagator 
$D_3(\bq,\omega)$, which, through Eqs.~\eqn{x3-dot} and \eqn{p3-dot}, 
implies that 
\[
D_3(\bq,\omega) = \fract{E_3-E_0}{\omega^2 - \omega_{3\bq}^2} \simeq - \fract
{E_3-E_0}{\omega_{3\bq}^2}
\simeq - \fract
{E_3-E_0}{\omega_{3\mathbf{0}}^2}\,,
\]
where we have furthermore neglected the momentum dependence.

In this approximation the mode $x_3$ simply induces a non-retarded electron-electron interaction, which,  
within RPA+Exchange, leads to a change of the  
charge and spin susceptibilities, 
\be
\chi_{c(s)}(\bq,\omega) 
\to \fract{\chi(\bq,\omega)}{1+\Gamma_{c(s)}(\bq)\,\chi(\bq,\omega)}
\;,\label{new-chi}
\ee 
where 
\be
\Gamma_{c(s)} \to \Gamma_{c(s)}(\bq) = 
\Gamma_{c(s)} - \fract{t^2\,u^2}{4T_0}\;\gamma_\bq\,,
\label{new-Gamma}
\ee
which also implies that the Landau parameters change into 
\be
F_0^{S(A)} \to -\mathcal{N}_*\,\Gamma_{c(s)}(\mathbf{0})\,.
\label{new-F}
\ee
The charge $F_0^{S}>0$ keeps its singularity $(1-u)^{-2}$, so that 
the charge compressibility still vanishes. On the contrary,
\be
F_0^A \underset{u\to 1}{\longrightarrow} \fract{t^2}{4T_0}\;
\mathcal{N}_*\,,\label{new-FA}
\ee 
so that the uniform spin susceptibility
\be
\chi = -\chi_s(\bq\to\mathbf{0},0) \underset{u\to 1}{\longrightarrow}
\fract{4T_0}{t^2} =  \fract{U_c}{2t^2}\,,
\label{new-spin-susceptibility}
\ee   
is now finite. Remarkably, this expression agrees with that obtained by DMFT\cite{ReviewDMFT}, although the numerical value of $U_c$ in DMFT is smaller than in the Gutzwiller approximation.\\
The quantum Hamiltonian \eqn{H-with 3} also allows calculating the optical conductivity. In the presence of 
a small transverse vector potential $A_{i\to j}(t)= -A_{j\to i}(t)$ the Hamiltonian acquires an additional term
\begin{align*}
\delta \mathcal{H}_*(t)  =& -i\fract{t_*}{\sqrt{z}}\,\sum_{<ij>}\,A_{i\to j}(t)
\Big(\bo{c}_i^\dagger \bo{c}_j^\dagga - H.c.\Big)X(i)\,X(j)\\
&+ \fract{t_*}{2\sqrt{z}}\,\sum_{<ij>}\,A_{i\to j}(t)^2
\Big(\bo{c}_i^\dagger \bo{c}_j^\dagga + H.c.\Big)X(i)\,X(j)\,. 
\end{align*}
The calculation of the optical conductivity is straightforward, and follows exactly that obtained within slave-bosons in Ref.~\onlinecite{Raimondi1993}. Besides the Drude peak that is obtained taking $X(i)=1$, and vanishes like $\sin^2\theta=1-u^2$ at the transition, the optical conductivity gets high-frequency contributions 
from the absorption spectrum of the mode $x_3$\cite{Raimondi1993}.

\section{Conclusions}
\label{Conclusions}
In this paper we have presented a quite simple method to calculate linear response functions within the Gutzwiller approximation,  including in a consistent way quantum fluctuations in the harmonic approximation. The calculation is straightforward and just requires a little more effort than the equilibrium one. In fact, besides the variational matrix $\hPhi_0$ that minimises the energy at equilibrium, and which can be regarded as the lowest energy eigenstate of a local Hamiltonian\cite{NicolaPRB2012,NicolaPRX}, see Eq.~\eqn{E-Phi}, one also needs all excited eigenstates and eigenvalues. In a model that involves $M$ correlated orbitals in each unit cell, this local Hamiltonian is defined in a Hilbert space of dimension $\binom{4M}{2M}$, and can be conveniently recast into the problem of an impurity with $M$ orbitals hybridised to a single bath site with the same number of orbitals, the coupled system being at half-filling\cite{NicolaPRX}. \\

As a check we have applied the method to the single-band Hubbard model at half-filling and recovered all known results\cite{VollhardtRMP,Jolicoeur1991,Raimondi1993,LorenzanaPRL,Lorenzana-PRB2003,LorenzanaPRB2004,SchiroFabrizioPRB,Sandri2012}. As a by-product, we also showed how to cure one flaw of the Gutzwiller approximation, i.e. the divergence of the uniform magnetic susceptibility approaching the Mott transition from the metal side. 

\section*{Acknowledgments}
This work has been supported by 
the European Union under H2020 Framework Programs, ERC Advanced Grant No. 692670 ``FIRSTORM''. 

\appendix

\section{The wavefunction renormalisation matrix $\hat{R}(i)$}
\label{app-1-1}
At equilibrium and in the natural basis, the constraint Eq.~\eqn{2} reads 
\ba
\Tr\Big(\hPhi_0(i)^\dagger\,\hPhi_0(i)\,\hat{d}^\dagger_{i\alpha}\,\hat{d}^\dagga_{i\beta}\Big) 
&=& \Tr\Big(\hat{P}^{(0)}_0(i) \, \hat{d}^\dagger_{i\alpha}\,\hat{d}^\dagga_{i\beta}\Big)\\
&&  = \delta_{\alpha\beta}\,
n^{(0)}_\alpha(i)\,,
\ea
where $\hat{P}_0(i)$ is the local probability distribution of the Slater determinant. Hereafter we shall drop for simplicity the site index $i$.  \\
We can always write $\hat{P}^{(0)}_0$ as the Boltzmann distribution of a non-interacting Hamiltonian 
\[
H = \sum_\alpha\,\epsilon_\alpha\,n_\alpha\,,
\]
where $f\big(\epsilon_\alpha\big) = n^{(0)}_\alpha$ 
is the Fermi distribution function. If $\hPhi$ is varied, also the probability distribution must vary in such a way as to preserve the constraint. This change will generally correspond to 
\[
H  \to H + \delta H
\,.
\]
Since $H$ must still be a one body Hamiltonian it follows that 
\[
d^\dagga_\alpha(\tau) = \text{e}^{\tau H}\;d^\dagga_\alpha\;\text{e}^{-\tau H} = 
\Big(\text{e}^{-\hat{H}\,\tau}\;\mathbf{d}^\dagga\Big)_\alpha
= \sum_\beta\, U_{\beta\alpha}(\tau)\,d^\dagga_\beta\, ,
\]
where $\hat{H}$ is the matrix representation of $H$ in the single-particle basis, so that $d^\dagga_\alpha(\tau)$ 
remains a combination of creation operators. 
Since 
$
\hat{U}(\tau_1)\;\hat{U}(\tau_2)=\hat{U}(\tau_1+\tau_2)$, 
it trivially holds that $\hat{U}(\tau)\,\hat{U}(-\tau)=1$ and 
\be
\hat{U}(\beta/2)\;\hat{U}(\beta/2)=\hat{U}(\beta)\,. \label{UvsU}
\ee
The local probability distribution   
\[
\hat{P}_0 = \fract{\text{e}^{-\beta \hat{H}}}{\Tr\Big(\text{e}^{-\beta \hat{H}}\Big)} \,,
\] 
so that    
\ba
&&\Tr\Big(\hat{P}_0\,\hat{d}^\dagga_\beta(\beta)\,\hat{d}^\dagger_\alpha\Big) =
\Tr\Big(\hat{P}_0\,\hat{d}^\dagger_\alpha\,\hat{d}^\dagga_\beta\Big) \equiv n_{\alpha\beta}\\
&&= 
\sum_\gamma\, U_{\gamma \beta}(\beta)\, \Tr\Big(\hat{P}_0\,d^\dagga_\gamma\,d^\dagger_\alpha\Big) 
= \sum_\gamma\, U_{\gamma \beta}(\beta)\,\Big(\delta_{\alpha\gamma} -n_{\alpha\gamma}\Big)\\
&&\qquad = U_{\alpha \beta}(\beta) - \sum_\gamma\, n_{\alpha\gamma}\;U_{\gamma\beta}(\beta)\,
\,,
\ea
namely 
\be
\hat{U}(\beta) = \Big(1 - \hat{n}\Big)^{-1}\;\hat{n} = 
 -1 + \Big(1 - \hat{n}\Big)^{-1}\,,\label{Uvsn}
\ee
which relates $\hat{U}(\beta)$ to $\hat{n}$. It also follows that  
\be
\hat{U}(-\beta) = \Big(1 - \hat{n}\Big)\;\hat{n}^{-1} = 
 \hat{n}^{-1}\;\Big(1 - \hat{n}\Big) = 
\hat{n}^{-1} - 1\,.\label{U-1vsn}
\ee\\
The renormalisation coefficients $R$ is obtained by solving for any 
$\alpha$ and $\gamma$ 
\bea
&&\Tr\Bigg(\sqrt{\hat{P}_0\;}\; \hPhi^\dagger\,\hat{c}^\dagger_\alpha\,\hPhi\; 
 \fract{1}{\;\sqrt{\hat{P}_0\;}\,}\; \hat{d}^\dagga_\gamma\Bigg) 
 \nonumber\\
&& \qquad\qquad =  \sum_\beta\Tr\Big(\hPhi^\dagger\,\hPhi\, \hat{d}^\dagger_\beta\,\hat{d}^\dagga_\gamma\Big)\;
R_{\alpha\beta}^*\,,\label{app-R}
\eea
where
\ba
 \fract{1}{\;\sqrt{\hat{P}_0\;}\;}\; \hat{d}^\dagga_\gamma\;\sqrt{\hat{P}_0\;} &=& 
\text{e}^{\,\beta \hat{H}/2}\; \hat{d}^\dagga_\gamma\;\text{e}^{-\beta \hat{H}/2} \\
&=& \hat{d}^\dagga_\gamma(\beta/2)
= \sum_\beta \, U_{\beta\gamma}(\beta/2)\, \hat{d}^\dagga_\delta
\,.
\ea
Therefore, once we define
\ba
Q_{\alpha\beta}^* &\equiv& \Tr\Big(\hPhi^\dagger\,\hat{c}^\dagger_\alpha\,\hPhi\,\hat{d}^\dagga_\beta\,\Big) \,,
\ea
then Eq.~\eqn{app-R} is equivalent to 
\[
\sum_\beta\, Q_{\alpha\beta}^*\; U_{\beta \gamma}\big(\beta/2\big)\; = 
\sum_\beta\, R_{\alpha\beta}^*\; n_{\beta\gamma}\,,
\]
or, in matrix form, and observing that $
\hat{n} = \hat{U}(\beta) - \hat{n}\;\hat{U}(\beta)$,
\ba
\hat{Q}^*\,\hat{U}(\beta/2) &=& \hat{R}^*\;\hat{n}= \hat{R}^*\;\Big(\hat{U}(\beta) - \hat{n}\;\hat{U}(\beta)\Big)\\ 
&& = \hat{R}^*\;\hat{U}(\beta) - \hat{Q}^*\;\hat{U}(3\beta/2) \, ,
\ea
so that, multiplying both sides on the right by $\hat{U}(-\beta)$ we finally get
\ba
\hat{R}^* &=& \hat{Q}^*\,\Big(\hat{U}(\beta/2) +\hat{U}(-\beta/2) \Big) \nonumber\\
&=& \hat{Q}^*\,\bigg(\sqrt{\hat{U}(\beta)\;} + 
\sqrt{\hat{U}(-\beta)\;}\,\bigg)\nonumber \\
&=& \hat{Q}^*\,\bigg(\sqrt{\fract{\hat{n}}{\;1-\hat{n}\;}\;} + 
\sqrt{\fract{\;1-\hat{n}\;}{\hat{n}}\;}\,\bigg) \nonumber\\
&& = 
\hat{Q}^*\;\bigg(\;\sqrt{\hat{n}\big(1-\hat{n}\big)\;}\;\bigg)^{-1}
\,.\label{Rvsall}
\ea
We denote as 
\[
\hat{S}^* = \bigg(\;\sqrt{\hat{n}\big(1-\hat{n}\big)\;}\;\bigg)^{-1}\;= \hat{S}^{\,\text{T}},
\]
since $\hat{S}=\hat{S}^\dagger$, so that 
\[
\hat{R}^* = \hat{Q}^*\,\hat{S}^*
\;\longrightarrow\; \hat{R}^\dagger = \hat{S}^\dagger\,\hat{Q}^\dagger = \hat{S}\,\hat{Q}^\dagger\,,
\] 
namely the desired result 
\be 
\hat{R} = \hat{Q}\,\hat{S}
\,.\label{R:QS}
\ee
One can rewrite 
\ba
4\,\hat{S}^{-2} &=& 4\hat{n}^\text{T}\,\Big(
1-\hat{n}^\text{T}\Big) = 1 -\Big(1-2\hat{n}^\text{T}\Big)^2
\equiv 1 - \hat{\Delta}^2\,,
\ea
where the matrix elements of $\hat{\Delta}$ are 
\bea
\Delta_{\alpha\beta} &=& \delta_{\alpha\beta} 
-2\Tr\Big(\hPhi^\dagger\,\hPhi\,\hat{d}^\dagger_\beta\,
\hat{d}^\dagga_\alpha\Big)\nonumber\\
&=& \Tr\bigg(\hPhi^\dagger\,\hPhi
\,\Big[
\hat{d}^\dagga_\alpha\,,\,\hat{d}^\dagger_\beta\Big]\bigg)\,.
\label{app-Delta}
\eea\\
At equilibrium 
\bea
\Delta^\z_{\alpha\beta} &=& \delta_{\alpha\beta}\,
\Big(1-2n^\z_\alpha\Big)\,,\label{app-Delta0}\\
S^\z_{\alpha\beta} &=& \delta_{\alpha\beta}/\sqrt{n^\z_\alpha\Big(1-n^\z_\alpha\Big)\;}
\equiv \delta_{\alpha\beta}\,S^\z_\alpha\;,\label{app-S0}
\eea
are diagonal, which allow an explicit 
evaluation of matrix derivatives. It follows that the equilibrium renormalisation matrix has elements 
\be
R^\z_{\alpha\beta} = \Tr\Big(\hPhi_0^\dagger\,\hat{c}^\dagga_\alpha\,\hPhi_0\,
\hat{d}^\dagger_\beta\Big)\;S^\z_\beta \equiv Q^\z_{\alpha\beta}\,S^\z_\beta\,.\label{app-R0}
\ee

\subsection{Derivatives of $\hat{R}$}
\label{app-1-2}
We write
\[
\hPhi = \sum_n\,\phi_n\,\hPhi_n\,,\qquad 
\hPhi^\dagger = \sum_n\,\phi_n^*\;\hPhi_n^\dagger\,,
\]
where $\hPhi_n$ is a basis set,
\[
\Tr\Big(\hPhi_n^\dagger\,\hPhi_m^\dagga\Big) = \delta_{nm}\,,
\] 
with $\hPhi_0$  the equilibrium solution. 
By inspection we realise that 
\ba
\fract{\partial R_{\alpha\beta}}{\partial \hPhi^\dagger} 
= \hat{\Gamma}_{\alpha\beta}\Big[\hPhi,\hPhi^\dagger\Big]\;\hPhi\,,
\ea 
where the tensor $\hat{\Gamma}_{\alpha\beta}\Big[\hPhi,\hPhi^\dagger\Big]$ is still functional of $\hPhi$ and $\hPhi^\dagger$.
Therefore 
\ba
\fract{\partial R_{\alpha\beta}}{\partial \phi_n^*}
= \Tr\bigg(\hPhi_n^\dagger\;
\hat{\Gamma}_{\alpha\beta}\Big[\hPhi,\hPhi^\dagger\Big]\;\hPhi\bigg)\,.
\ea
The equilibrium value is obtained by setting $\phi_n=\delta_{n0}$. \\
In particular, exploiting the fact that $\hat{S}$ is diagonal at equilibrium, the first order derivatives 
evaluated at equilibrium read explicitly 
\begin{align}
\fract{\partial R_{\alpha\beta}}{\partial \phi_n^*}
&= \fract{\partial Q_{\alpha\beta}}{\partial \phi_n^*}\,S^\z_\beta 
+ \sum_\gamma\,Q^\z_{\alpha\gamma}\,S^\z_\gamma\, F_{\gamma\beta}\,
\fract{\partial \Delta_{\gamma\beta}}{\partial \phi_n^*}\,,\label{app-R-1}\\
\fract{\partial R_{\alpha\beta}}{\partial \phi_n}
&= \fract{\partial Q_{\alpha\beta}}{\partial \phi_n}\,S^\z_\beta 
+ \sum_\gamma\,Q^\z_{\alpha\gamma}\,S^\z_\gamma\, F_{\gamma\beta}\,
\fract{\partial \Delta_{\gamma\beta}}{\partial \phi_n}\,,\label{app-R-1-*}
\end{align}
while the second derivative, still calculated at equilibrium, is 
\bea
\fract{\partial^2 R_{\alpha\beta}}{\partial \phi_n^*\partial \phi_m}
&=& \Tr\bigg(\hPhi_n^\dagger\;
\hat{\Gamma}_{\alpha\beta}\Big[\hPhi_0,\hPhi_0^\dagger\Big]\;\hPhi_m\bigg) \nonumber\\
&& + 
\Tr\Bigg(\hPhi_n^\dagger\;
\fract{\partial \hat{\Gamma}_{\alpha\beta}\Big[\hPhi,\hPhi^\dagger\Big]}{\partial \phi_m}_{\big|0}\;\hPhi_0\Bigg)\label{R-second-derivative}\\
&\equiv& \left(\fract{\partial^2 R_{\alpha\beta}}{\partial \phi_n^*\partial \phi_m}\right)_1 + 
\left(\fract{\partial^2 R_{\alpha\beta}}{\partial \phi_n^*\partial \phi_m}\right)_2
\,,\nonumber
\eea
where 
\bw 
\bea
\left(\fract{\partial^2 R_{\alpha\beta}}{\partial \phi_n^*\partial \phi_m}\right)_1 \!\!&=& 
\sum_\gamma\,\Bigg[
\fract{\partial^2 Q_{\alpha\beta}}{\partial \phi_n^*\partial \phi_m} \;
S^\z_{\beta} + 
Q^\z_{\alpha\gamma}\,F_{\gamma\beta}\;\fract{\partial^2 \Delta_{\gamma \beta}}{\partial \phi_n^*\partial \phi_m}\;\Bigg]\,,\label{important-1}\\
\left(\fract{\partial^2 R_{\alpha\beta}}{\partial \phi_n^*\partial \phi_m}\right)_2 &=&
\sum_\gamma\,\Bigg[
\fract{\partial Q_{\alpha\gamma}}{\partial \phi_n^*}\,F_{\gamma\beta}
\;\fract{\partial \Delta_{\gamma\beta}}{\partial \phi_m}
+ \fract{\partial Q_{\alpha\gamma}}{\partial \phi_m}\,F_{\gamma\beta}
\;\fract{\partial \Delta_{\gamma\beta}}{\partial \phi_n^*}
+ Q^\z_{\alpha\gamma}\,
\left(\fract{\partial^2 S_{\gamma\beta}}
{\partial \phi_n^*\partial\phi_m}\right)_2
\;\Bigg]
\,.\;\;\;\; \qquad \label{important-2}
\eea
\ew
The terms that appear in the above equations are 
\begin{align*}
\fract{\partial Q_{\alpha\beta}}{\partial \phi_n^*}
&= \Tr\Big(\hPhi_n^\dagger\,\hat{c}^\dagga_\alpha\,\hPhi_0\,\hat{d}^\dagger_\beta\Big)\,,\\
\fract{\partial Q_{\alpha\beta}}{\partial \phi_n}
&= \Tr\Big(\hPhi_0^\dagger\,\hat{c}^\dagga_\alpha\,\hPhi_n\,\hat{d}^\dagger_\beta\Big)\,,\\
\fract{\partial \Delta_{\alpha\beta}}{\partial \phi_n^*}
&=\Tr\bigg(\hPhi_n^\dagger\,\hPhi_0\,\Big[\hat{d}^\dagga_{\alpha}\,,\,
\hat{d}^\dagger_{\beta}\Big]\bigg)\,,\\
\fract{\partial \Delta_{\alpha\beta}}{\partial \phi_n}
&=\Tr\bigg(\hPhi_0^\dagger\,\hPhi_n\,\Big[\hat{d}^\dagga_{\alpha}\,,\,
\hat{d}^\dagger_{\beta}\Big]\bigg)\,,\\
F_{\alpha\beta} &= \fract{1}{2}\;\fract{\left(S^\z_\alpha\,S^\z_\beta\right)^2}
{S^\z_\alpha+S^\z_\beta}\;\left(1-n^\z_\alpha-n^\z_\beta\right)\,,\\
\fract{\partial^2 Q_{\alpha\beta}}
{\partial \phi_n^*\partial\phi_m} &= 
\Tr\Big(\hPhi_n^\dagger\,\hat{c}^\dagga_{\alpha}\,\hPhi_m\,
\hat{d}^\dagger_{\beta}\Big)\,,\\
\fract{\partial^2 \Delta_{\alpha\beta}(i)}
{\partial \phi_n^*\partial\phi_m} &= 
\Tr\bigg(\hPhi_n^\dagger\,\hPhi_m\,\Big[\hat{d}^\dagga_{\alpha}\,,\,
\hat{d}^\dagger_{\beta}\Big]\bigg)\,,
\end{align*}
and, lastly, 
\bw
\ba
\left(\fract{\partial^2 S_{\alpha\beta}}
{\partial \phi_n^*\partial\phi_m}\right)_2\!\! \!&=& \!
\fract{\Big(S^\z_\alpha\,S^\z_\beta\Big)^2}{S^\z_\alpha+S^\z_\beta}\!\sum_\gamma\!
\Bigg[\fract{\partial \Delta_{\alpha\gamma}}{\partial \phi_n^*}\,
\fract{\partial \Delta_{\gamma\beta}}{\partial \phi_m}\!+\!
\fract{\partial \Delta_{\alpha\gamma}}{\partial \phi_m}\,
\fract{\partial \Delta_{\gamma\beta}}{\partial \phi_n^*}\Bigg]\!\!
\left(\frac{1}{4} + F_{\alpha\gamma}\,F_{\gamma\beta}
\fract{S^\z_\alpha S^\z_\gamma+ S^\z_\gamma S^\z_\beta+ S^\z_\beta S^\z_\alpha}
{\Big(S^\z_\alpha S^\z_\gamma S^\z_\beta\Big)^2}\right).
\ea
\ew 
In addition 
\bea
\fract{\partial^2 R_{\alpha\beta}}{\partial \phi_n^*\partial \phi_m^*}
&=& 
\left(\fract{\partial^2 R_{\alpha\beta}}{\partial \phi_n^*\partial \phi_m^*}\right)_2\,,
\label{R-second-derivative-1}\\
\fract{\partial^2 R_{\alpha\beta}}{\partial \phi_n\partial \phi_m}
&=& 
\left(\fract{\partial^2 R_{\alpha\beta}}{\partial \phi_n\partial \phi_m}\right)_2\,,
\label{R-second-derivative-2}
\eea
where the right hand sides are obtained straightforwardly through Eq.~\eqn{important-2}. 
The above derivatives calculated at the equilibrium solution allow calculating the Taylor expansion of $\hat{R}$. In particular, through equations \eqn{app-R-1} and \eqn{app-R-1-*}, the first order expansion 
is
\be
\hat{R}^{(1)} = 
\sum_{n}\,\left[\,\phi_n^*\,\fract{\partial R_{\alpha\beta}}{\partial \phi_n^*}
+ \phi_n\,\fract{\partial R_{\alpha\beta}}{\partial \phi_n}\,\right]\,,\label{app-R-1-final}
\ee
while the second order expansion mentioned in Eq.~\eqn{R-2-12}, is 
\be
\hat{R}^{(2)} = \hat{R}_1^{(2)} + \hat{R}_1^{(2)}\,,\label{app-R-2-12}
\ee
where, explicitly, 
\be
\hat{R}_1^{(2)} = \sum_{nm}\,\phi_n^*\,\phi_m\;
\left(\fract{\partial^2 \hat{R}}{\partial \phi_n^*\partial \phi_m}\right)_1\,,\label{app-R-2-1}
\ee
and 
\bw
\bea
\hat{R}_2^{(2)} &=& \fract{1}{2}\,\sum_{nm}\,\left[\,2\,\phi_n^*\,\phi_m\;
\left(\fract{\partial^2 \hat{R}}{\partial \phi_n^*\partial \phi_m}\right)_2 
+ \phi_n^*\,\phi_m^*\;
\left(\fract{\partial^2 \hat{R}}{\partial \phi_n^*\partial \phi_m^*}\right)_2
+ \phi_n\,\phi_m\;
\left(\fract{\partial^2 \hat{R}}{\partial \phi_n\partial \phi_m}\right)_2\;\right]\,.
\label{app-R-2-2}
\eea
\ew


\end{document}